\documentclass[reqno,a4paper]{amsart}
\usepackage[T1]{fontenc}
\usepackage{comment,amssymb,latexsym,upref,enumerate,fouridx}
\usepackage[normalem]{ulem}
\usepackage{dutchcal}
\usepackage{mathrsfs,xcolor}
\usepackage{graphicx}
\usepackage{subfigure}
\usepackage{placeins}
\usepackage{centernot}
\usepackage[colorlinks,linkcolor=blue,citecolor=blue]{hyperref}

\allowdisplaybreaks
\numberwithin{equation}{section}

\theoremstyle{plain}
\newtheorem{thm}{Theorem}[section]

\theoremstyle{definition}

\theoremstyle{remark}
\newtheorem{rem}[thm]{Remark}

\newcommand*{\dd}{\mathrm{d}}

\input Tdef.def

\newcounter{mnotecount}[section]
\let\oldmarginpar\marginpar
\renewcommand\marginpar[1]{\-\oldmarginpar[\raggedleft\footnotesize #1]%
 {\raggedright\footnotesize #1}}

\graphicspath{{./}{./Images/Convergence_Plots/}{./Images/Asymptotic_Behaviour/}{./Images/Density_Gradient/}{./Images/}{./Images/RicciScalar/}{./Images/Spikes/}}

\begin{document}
\title[Past instability of FLRW solutions of the Einstein-Euler-scalar field equations for linear equations of state $p=K\rho$ with $0 \leq K<1/3$]{Past instability of FLRW solutions of the Einstein-Euler-scalar field equations for linear equations of state $p=K\rho$ with $0 \leq K<1/3$} 

\author[F.~Beyer]{Florian Beyer}
\address{Dept of Mathematics and Statistics\\
730 Cumberland St\\
University of Otago, Dunedin 9016\\ New Zealand}
\email{florian.beyer@otago.ac.nz}

\author[E.~Marshall]{Elliot Marshall}
\address{School of Mathematics\\
9 Rainforest Walk\\
Monash University, VIC 3800\\ Australia}
\email{elliot.marshall@monash.edu}

\author[T.A.~Oliynyk]{Todd A.~Oliynyk}
\address{School of Mathematics\\
9 Rainforest Walk\\
Monash University, VIC 3800\\ Australia}
\email{todd.oliynyk@monash.edu}

\begin{abstract} 
Using numerical methods, we examine, under a Gowdy symmetry assumption, the dynamics of nonlinearly perturbed FLRW fluid solutions of the Einstein-Euler-scalar field equations in the contracting direction for linear equations of state $p = K\rho$ and sound speeds $0\leq K<1/3$.
This article builds upon the numerical work from \cite{BMO:2023} in which perturbations of FLRW solutions to the Einstein-Euler equations with positive cosmological constant in the expanding time direction were studied. The numerical results presented here confirm that the instabilities observed in \cite{BMO:2023,MarshallOliynyk:2022} for $1/3<K<1$, first conjectured to occur in the expanding direction by Rendall in \cite{Rendall:2004}, are also present in the contracting direction over the complementary parameter range $0\leq K<1/3$. Our numerical solutions show
that the fractional density gradient 
of the nonlinear perturbations develop steep gradients near a finite number of spatial points and become unbounded towards the big bang. This behaviour, and in particular the characteristic profile of the fractional density gradient near the big bang, is strikingly similar to what was observed in the expanding direction near timelike infinity in the article \cite{BMO:2023}.
\end{abstract}

\maketitle

\section{Introduction\label{intro}}
Perturbed Friedmann-Lemaître-Robertson-Walker (FLRW) spacetimes form the basis of modern cosmology and play a distinguished role in General Relativity. 
In particular, much research has been devoted to understanding the dynamical behaviour of these spacetimes near big bang  singularities. Due to the Hawking-Penrose singularity theorems \cite{HawkingEllis:1973}, it is known that cosmological spacetimes without any symmetries are
geodesically incomplete to the past (contracting direction) for a large class of matter models. However, it has only been recently established that the past geodesic incompleteness of perturbed Einstein-scalar field FLRW spacetimes, possibly coupled with other matter fields, is due to the formation of quiescent, spacelike big bang singularities where the curvature becomes unbounded \cite{BeyerOliynyk:2024,FajmanUrban:2022,Fournodavlos_et_al:2023,Groeniger_et_al:2023,RodnianskiSpeck:2018a, RodnianskiSpeck:2018b,RodnianskiSpeck:2018c,Speck:2018}, which is referred to as \textit{FLRW big bang stability}. More generally, the stability of big bang singularities in the Kasner family of solutions to the  Einstein-scalar field equations has been established in  \cite{Fournodavlos_et_al:2023,Groeniger_et_al:2023} for the expected range of quiescent Kasner exponents.

Scalar fields are thought to play an important role during the evolution of the early universe \cite{Bardeen_et_al:1983,Guth:81}. So, 
while quiescent big bang singularities are not expected to be generic for most matter models, and indeed this is the view put forward by the BKL conjecture \cite{belinskii1970,lifshitz1963}, the presence of scalar fields in the early universe imply that quiescent big bang singularities are physically relevant because nonlinear perturbations of FLRW solutions to the Einstein-scalar field system terminate in the past at such singularities. In addition to scalar fields, fluids, and in particular, radiation fluids, must also be considered as matter models of the early universe. First steps in this direction were taken in the articles \cite{RodnianskiSpeck:2018c} ($K=1$ and no scalar field) and \cite{BeyerOliynyk:2023} ($1/3<K<1$), where the stability of FLRW big bang singularities for solutions of the Einstein-Euler-scalar field equations with linear equations of state $p=K\rho$ was established. While these stability results go some way towards understanding the behaviour of fluids near FLRW big bang singularities, they do not apply to radiation fluids $K=1/3$ or to the case $K<1/3$. As discussed in \cite{BeyerOliynyk:2023}, see also Section~\ref{sec:asympbehav}, it is expected that over the range of sounds speeds $0\leq K\leq 1/3$ fluids will behave significantly differently compared to $1/3<K<1$.

The aim of this article is to numerically investigate the behaviour towards the past of nonlinear perturbations of FLRW solutions to the Einstein-Euler-scalar field equations for the range of sound speeds $0\leq K\leq 1/3$. Of particular interest is to resolve the behaviour of the gravitational and matter fields near big bang singularities that form in these perturbed solutions.  In order to simplify the problem, we restrict our attention to spatial $\Tbb^3$-toplogies and numerically solve the Einstein-Euler-scalar field equations under a Gowdy symmetry assumption (see Section~\ref{sec:EinsteinEulerScalarderivationGowdy}). The advantage of considering Gowdy spacetimes is that the presence of two Killing fields allows us to reduce the Einstein-Euler-scalar field equations to a $(1+1)$-dimensional problem with periodic boundary conditions. This type of simplification has been exploited both analytically and numerically many times in previous studies of the Einstein equations \cite{ames2017,amorim2009,berger:1993,BeyerHennig:2012,beyer2010b,beyer2017,beyer2020b,chrusciel1990,isenberg1990,kichenassamy1998,LeFlochRendall:2011,rendall2000,ringstrom2009a}.

The numerical simulations we perform reveal that nonlinear Gowdy-symmetric perturbations of FLRW solutions to the Einstein-Euler-scalar field equations display the following behaviour: 
\begin{enumerate}[(a)]
\item For all sounds speeds $K=c_s^2\in [0,1]$ and all sufficiently small perturbations of FLRW initial data, a spacelike big bang singularity forms in areal coordinates (see \eqref{eqn:gowdymetric}) at $\tb=0$ and the Ricci scalar blows up there.
\item For $K\in [0,1/3)$ and initial data that is sufficiently close to FLRW initial data and for which the spatial velocity vanishes somewhere on the initial hypersurface,  the \textit{fractional density gradient} $\frac{\del_{\theta}\rho}{\rho}$ develops steep gradients and
blows up at finitely many spatial points at $\tb=0$. These blow up points coincide with the vanishing of the spatial velocity at $\tb=0$. We refer to the sharp features that form near $\tb=0$ in the fractional density gradient as \textit{spikes}.
Moreover, at the spatial points where the spatial fluid velocity vanishes at $\tb=0$, the fluid behaves asymptotically as $\tb\searrow 0$ like an \textit{orthogonal} fluid, while away from these points it behaves asymptotically like a \textit{tilted} fluid.
\item  At $K=1/3$ and for initial data that is sufficiently close to FLRW initial data, we observe no blow-up of the fractional density gradient and it appears that all the (suitably renormalised) fluid and gravitational variables are converging as $\tb\searrow 0$. However, as the blow-up of the fractional density that occurs for $0\leq K <1/3$ takes longer and longer to set in as $K$ approaches $1/3$, it could be the case that the perturbations are also unstable for $K=1/3$ and we are not observing it numerically because we are simply not evolving long enough to see the instability. 
\item  For $K\in (1/3,1]$ and initial data that is sufficiently close to FLRW initial data, all of the (suitably renormalised) gravitational and matter variables converge at $\tb\searrow 0$ monotonically to limits in accordance with the stability results \cite{BeyerOliynyk:2023,RodnianskiSpeck:2018c}.
\item  For initial data that is sufficiently far away from FLRW initial data and $K\in [0,1]$, spikes form in both fluid and metric functions. Intriguingly, the spikes in the fractional density gradient form first and are followed by gravitational spikes that develop in nearly the same location.
\end{enumerate}
These results can be understood by a simple heuristic argument. First suppose that, as the big bang is approached, the dynamics of the Einstein-Euler-scalar field system can be approximated by solutions of the Euler equations on a fixed FLRW-scalar field background and that spatial derivatives are negligible. Then, as we show in Section~\ref{sec:asympbehav}, the relation 
\begin{equation}  
    \label{eq:specialimplsolfluid.Intro}
   \beta\bigl(1-\beta^2\bigr)^{-(1-K)/2} = c\, \bar t^{(3 K-1)/2}
\end{equation}
holds at each spatial point where $c$ is a constant and the fluid velocity field is determined in terms of $\beta$ and a natural orthonormal frame $\{e_0,e_1\}$ by
\begin{equation*}
    v=\frac {e_0+\beta e_1}{\sqrt{1-\beta^2}}.
\end{equation*} 
The function $\beta$, which represents the spatial part of the fluid velocity, takes values in $(-1,1)$. In particular,
\eqref{eq:specialimplsolfluid.Intro} implies that the asymptotic behaviour of the fluid at each spatial point is determined by the constants $c$ and $K$. This leads to the following classification of the asymptotic behaviour:
\begin{itemize}
    \item Orthogonal fluid: $c=0$, $K\in [0,1]$.
    \item Asymptotically orthogonal fluid: $c\not=0$, $K\in (1/3,1]$.
    \item Tilted fluid: $c\not=0$, $K=1/3$.
    \item Asymptotically extremely tilted fluid\footnote{A fluid is described as having an `extreme tilt' if the leading order behaviour of the fluid, as the singularity is approached, is a null vector.}: $c\not=0$, $K\in [0,1/3)$.
\end{itemize}
The instability described above for $K\in [0,1/3)$ is thus driven by the dramatically different behaviour of the orthogonal and asymptotically extremely tilted fluids. 

Interestingly \eqref{eq:specialimplsolfluid.Intro} also reveals that the opposite dichotomy occurs in the future (expanding) direction corresponding to the limit $\bar t\rightarrow\infty$ interchanging the sound speed parameter $K$ range. In fact, FLRW fluid stability for $0\leq K\leq 1/3$ has been established in \cite{Friedrich:2017,HadzicSpeck:2015,LiuOliynyk:2018b,LiuOliynyk:2018a,LubbeKroon:2013,Oliynyk:CMP_2016,RodnianskiSpeck:2013,Speck:2012} and instability for $1/3<K\leq 1$ in \cite{BMO:2023,Fournodavlos:2022}. The FLRW fluid instabilities that were observed numerically in Gowdy-symmetric solutions for $1/3<K<1$ in \cite{BMO:2023} were predicted by Rendall \cite{Rendall:2004} and are driven by the blow-up of the fractional density gradient $\frac{\del_{\theta}\rho}{\rho}$. Indeed, it was observed in \cite{BMO:2023} that for all $K\in (1/3,1)$ and all choices of initial data sufficiently close to FLRW initial data, the fractional density gradient  develops steep gradients and
blows up at finitely many spatial points at future timelike infinity.\footnote{See \cite{Oliynyk:2023} for a rigorous analysis of this instability in the simplified setting where coupling to Einstein's equations is ignored.} It is also interesting to note that the blow-up profiles of the fractional density gradient near future timelike infinity observed in \cite{BMO:2023} are remarkably similar to the blow-up profiles of the fractional density gradient near the big bang singularity at $\tb=0$ in the numerical simulations presented here.

\subsection{Prior and related results:} 
The fluid instability described above in points (b) and (e) is referred to as a \textit{tilt-instability} in \cite{ColeyLim:2013}. In that article and also \cite{ColeyLim:2012,ColeyLim:2015}, the authors construct analytic and numerical solutions of the Einstein-Euler equations with a $G_{2}$ symmetry\footnote{$G_{2}$ models include Gowdy spacetimes as a special case, see \cite{LAGP:2009}.} that exhibit spikes in both the fluid and gravitational fields and are clearly related to what we observe numerically in this article, see point (e) above. In contrast, the stability dichotomy, see points (b)-(d) above, that we observe in this article for sufficiently small perturbations of FLRW solutions of the Einstein-Euler-scalar field equations is new, as is the clear characterisation of the fluid instability for $0\leq K<1/3$ as blow-up of the fractional density gradient $\frac{\del_{\theta}\rho}{\rho}$ at the big bang singularity located at $\tb=0$ and the identification of the fluid spikes as large gradients that develop in the fractional density gradient near $\tb=0$. It would be interesting to understand if a similar behaviour occurs for small perturbations of the Kasner family of solutions to the Einstein-Euler-scalar fields equations for all exponents that lie in the quiescent range.

\subsection{Overview:} 
The article is organised as follows: the derivation of a first order formulation of the Gowdy-symmetric Einstein-Euler-scalar field equations that is suitable for numerical implementation and constructing solutions globally to the future is carried out in Section \ref{sec:EinsteinEulerScalarderivation}. In Section \ref{sec:FLRWsoln}, we derive the FLRW background solutions that we perturb and in Section \ref{sec:NumericalResults} we discuss our numerical setup and results for small perturbations of the FLRW solution. Finally, in Section \ref{sec:BigDataSpikes} we investigate large perturbations and the interactions between gravitational spikes and spikes in the fractional density gradient.

\section{Einstein-Euler-scalar field Equations}
\label{sec:EinsteinEulerScalarderivation}

\subsection{Einstein-Euler-scalar field equations with Gowdy symmetry}
\label{sec:EinsteinEulerScalarderivationGowdy}

The Einstein-Euler-scalar field equations\footnote{Our indexing conventions are as follows: lower case Latin letters, e.g. $i,j,k$,
will label spacetime coordinate indices that run from $0$ to $3$ while upper case Latin letters, e.g. $I,J,K$, will label spatial coordinate indices that run from
$1$ to $3$.} for a perfect fluid and minimally-coupled scalar field are given by\footnote{Here, we use units where $c=1$ and $G=\frac{1}{8\pi}$.} 
\begin{align}
\label{eqn:Einstein1}
    G_{ij} &=T^{\text{fl}}_{ij}+T^{\phi}_{ij}, \\
\nabla^{i}T^{\text{fl}}_{ij} &= 0, \label{eqn:Fluid_Tij_divergence}\\
\nabla^{i}T^{\phi}_{ij} &= 0,
   \label{eqn:Tij_divergence}
\end{align}

where
\begin{align*}
    T^{\phi}_{ij} = \nabla_{i}\phi\nabla_{j}\phi - \frac{1}{2}g_{ij}\nabla^{a}\phi\nabla_{a}\phi 
\end{align*}
is the scalar field stress-energy tensor and
\begin{align*}
    T^{\text{fl}}_{ij} &= (\rho+p)v_{i}v_{j}+pg_{ij} 
\end{align*}
is the perfect fluid stress-energy tensor. Here, $v_{i}$ is the fluid four-velocity normalised by $v_{i}v^{i}=-1$, and we assume that the fluid's proper energy density, $\rho$, and pressure, $p$, are related via the linear equation of state 
\begin{align*}
    p = K\rho,
\end{align*}
where the constant parameter $K\geq 0$ is the square of the sound speed. In the following, we assume that $0\leq K \leq 1$ so that the speed of sound is less than or equal to the speed of light.

As discussed in the introduction, we restrict our attention to solutions of the Einstein-Euler-scalar field equations with a Gowdy symmetry \cite{chrusciel1990,gowdy1974} by
considering Gowdy metrics in areal coordinates on $\Rbb_{>0} \times \mathbb{T}^{3}$ of the form
\begin{align}
\label{eqn:gowdymetric}
    g = e^{2(\eta-U)}(-e^{2\bar{\alpha}} d\bar{t} \otimes d\bar{t}+ d\theta \otimes d\theta)+e^{2U}(dy+Adz)\otimes (dy+Adz)+e^{-2U}\bar{t}^{2}dz \otimes dz.
\end{align}
Here, the functions $\eta$, $U$, $\bar{\alpha}$, and $A$ depend only on $(\bar{t},\theta)\in \Rbb_{>0}\times \Rbb$ and are $2\pi$-periodic in $\theta$. Since our spatial slices are $\mathbb{T}^{3}$, the metric is periodic and compact in $y$ and $z$ as well, however, these coordinates play no role due to the symmetry condition. For an in-depth discussion of spacetimes with $U(1)\times U(1)$ symmetry and compact spatial slices see \cite{{chrusciel1990}}. Following previous numerical studies of the initial singularity in Gowdy symmetry \cite{berger1998c,Berger:2001,berger:1993}, we introduce a new time $t$ and metric function $\alpha$ via 
\begin{equation}
    \label{eq:timetransformation}
    \bar{t} = e^{-t}, \;\; \bar{\alpha} = \alpha +t,
\end{equation}
which allows us to express the Gowdy metric \eqref{eqn:gowdymetric} as
\begin{align}
    g = e^{2(\eta-U)}(-e^{2\alpha} dt \otimes dt+ d\theta \otimes d\theta)+e^{2U}(dy+Adz)\otimes (dy+Adz)+e^{-2U-2t}dz \otimes dz, \label{eqn:gowdymetricA}
\end{align}
where the big bang singularity is now located at $t = \infty$. We are only interested in solutions in the contracting direction, i.e.\ towards the past, and consequently, we consider time intervals of the form $t\in [t_0,\infty)$ for some $t_0\in\Rbb$.

Next, we turn to expressing the Einstein-Euler-scalar field system \eqref{eqn:Einstein1}-\eqref{eqn:Tij_divergence} in a Gowdy-symmetric form suitable for numerical implementation.
This involves expressing the Einstein and scalar field equations in first order form and choosing appropriate variables to formulate the Euler equations. The details of the derivation are presented in the following three sections. 

\subsection{A First Order Formulation of the Einstein Equations}

\label{sec:EinsteinEquations}
In Gowdy symmetry, the fluid four-velocity only has two non-zero components\footnote{This follows from choosing coordinates where the two Killing vectors are given by $\del_{y}$ and $\del_{z}$, see \cite{LeFlochRendall:2011}.} and can be expressed as
\begin{equation} \label{v-Gowdy}
v=v_0 dt  + v_1 d\theta,
\end{equation}
where the functions $v_0$ and $v_1$ depend on $(t,\theta)\in \Rbb\times \Rbb$ and are $2\pi$-periodic in $\theta$. Due to the normalisation $v_i v^i =-1$, only one of these functions are independent and we take $v_1$ as our primary fluid velocity variable. Furthermore, the scalar field in Gowdy symmetry also depends on $(t,\theta)\in \Rbb\times \Rbb$ and is $2\pi$-periodic in $\theta$.
With these choices, the non-zero components of the total stress-energy tensor
\begin{equation*}
    T_{ij} = T^{\text{fl}}_{ij}+T^{\phi}_{ij}
\end{equation*}
are given by
\begin{align*}
T_{00} &=-e^{2(\alpha+\eta-U)}K\rho+(1+K)\rho v_{0}^{2}+(\del_{t}\phi)^{2}+\frac{1}{2}e^{2(\alpha+\eta-U)}\Big(e^{2(U-\eta)}(\del_{\theta}\phi)^{2}-e^{-2(\alpha+\eta-U)}(\del_{t}\phi)^{2}\Big), \\
T_{01} &= (1+K)\rho v_{0}v_{1}+\del_{t}\phi\del_{\theta}\phi, \\
T_{11} &= e^{2(\eta-U)}K\rho+(1+K)\rho v_{1}^{2}+(\del_{\theta}\phi)^{2}-\frac{1}{2}e^{2(\eta-U)}\Big(e^{2(U-\eta)}(\del_{\theta}\phi)^{2}-e^{-2(\alpha+\eta-U)}(\del_{t}\phi)^{2}\Big), \\ 
T_{22} &= e^{2U}K\rho-\frac{1}{2}e^{2U}\Big(e^{2(U-\eta)}(\del_{\theta}\phi)^{2}-e^{-2(-U+\alpha+\eta)}(\del_{t}\phi)^{2}\Big), \\
T_{23} &= e^{2U}K\rho A-\frac{1}{2}e^{2U}A\Big(e^{2(U-\eta)}(\del_{\theta}\phi)^{2}-e^{-2(\alpha+\eta-U)}(\del_{t}\phi)^{2}\Big), \\ 
T_{33} &= (e^{2U}A^{2}+e^{-2U-2t})\Bigg(K\rho -\frac{1}{2}\Big(e^{2(U-\eta)}(\del_{\theta}\phi)^{2}-e^{-2(\alpha+\eta-U)}(\del_{t}\phi)^{2}\Big)\Bigg).
\end{align*}
Using these  expressions and the Gowdy metric \eqref{eqn:gowdymetricA}, 
a straightforward calculation shows that the Einstein equation \eqref{eqn:Einstein1} in Gowdy symmetry consists of following three wave equations
\begin{align}
\label{eqn:Awave}
\del_{tt}A &= e^{2\alpha}\Big(\del_{\theta}A(4\del_{\theta}U + \del_{\theta}\alpha) + \del_{\theta\theta}A\Big) + \del_{t}A(-1-4\del_{t}U+\del_{t}\alpha), \\
\label{eqn:Uwave}
\del_{tt}U &= \frac{1}{2} + \frac{1}{2}e^{2t+4U}\Big((\del_{t}A)^{2} - e^{2\alpha}(\del_{\theta}A)^{2}\Big) + e^{2\alpha}\del_{\theta}U\del_{\theta}\alpha + e^{2\alpha}\del_{\theta\theta}U + \del_{t}U + \frac{1}{2}\del_{t}\alpha + \del_{t}U\del_{t}\alpha, \\
\label{eqn:Etawave}
\del_{tt}\eta &= -\frac{1}{4}e^{-2U}\Big(4e^{2\alpha+2\eta}K\rho + e^{2t+6U+2\alpha}(\del_{\theta}A)^{2} - 4e^{2U+2\alpha}(\del_{\theta}U)^{2} - 4e^{2U+2\alpha}(\del_{\theta}\alpha)^{2} \nonumber \\
&- 4e^{2U+2\alpha}\del_{\theta}\alpha\del_{\theta}\eta - 2e^{2U+2\alpha}(\del_{\theta}\phi)^{2} - 4e^{2U +2\alpha}\del_{\theta\theta}\alpha \nonumber \\
&- 4e^{2U}(\del_{t}U)^{2} - 4e^{2U}\del_{t}\alpha\del_{t}\eta + 2e^{2U}(\del_{t}\phi)^{2}\Big),
\end{align}
and three first order equations
\begin{align}
\label{eqn:alphaevo_1}
\del_{t}\alpha &= -1 -2e^{2(\alpha+\eta-U)}K\rho + (1+K)\rho v_{0}^{2} -e^{2\alpha}(1+K)\rho v_{1}^{2}, \\
\label{eqn:Hamiltonian_constraint1}
\del_{t}\eta &= -\frac{1}{4}e^{-2U}\Big(-4e^{2(\alpha+\eta)}K\rho + 4e^{2U}\rho v_{0}^{2} + 4e^{2U}K\rho v_{0}^{2} + e^{2t +6U +2\alpha}(\del_{\theta}A)^{2} + 4e^{2(U+\alpha)}(\del_{\theta}U)^{2} \nonumber \\
&+ 2e^{2U+2\alpha}(\del_{\theta}\phi)^{2} + e^{2t+6U}(\del_{t}A)^{2} + 4e^{2U}(\del_{t}U)^{2} + 2e^{2U}(\del_{t}\phi)^{2}\Big), \\
\label{eqn:Momentum_constraint1}
\del_{\theta}\eta &= \frac{1}{2}\Big(-2\rho v_{0} v_{1} -2K\rho v_{0}v_{1} - 2\del_{\theta}\alpha - e^{2t+4U}\del_{\theta}A\del_{t}A - 4\del_{\theta}U\del_{t}U - 2\del_{\theta}\phi\del_{t}\phi\Big),
\end{align}
where we note that \eqref{eqn:Hamiltonian_constraint1} and \eqref{eqn:Momentum_constraint1} are the Hamiltonian and momentum constraints, respectively. 

Either of \eqref{eqn:Etawave} or \eqref{eqn:Hamiltonian_constraint1} can be used to evolve the metric variable $\eta$. Following our previous numerical study of the Gowdy-symmetric Einstein-Euler equations in the expanding direction \cite{BMO:2023}, we use \eqref{eqn:Hamiltonian_constraint1} to evolve $\eta$. This choice, as discussed in \cite{BMO:2023}, has the benefit of enforcing the Hamiltonian constraint\footnote{The importance of enforcing the Hamiltonian constraint for numerical simulations is further discussed in \cite{Berger:2006}.} at every time step and involves solving a first order equation for $\eta$ rather than a second order one. Moreover, because we use \eqref{eqn:Hamiltonian_constraint1} to evolve $\eta$, we can view \eqref{eqn:Etawave} as a constraint equation that can be used to verify our numerical results.

Next, introducing the first order variables
\begin{align}
\label{eqn:firstordervariablesA_U}
A_{0} &= \del_{t}A, \quad A_{1} = e^{\alpha}\del_{\theta}A, \quad U_{0} = \del_{t}U, \quad U_{1} = e^{\alpha}\del_{\theta}U,
\end{align}
we can express the wave equations \eqref{eqn:Awave}-\eqref{eqn:Uwave} for $A$ and $U$ in first order form as
\begin{align}
\label{eqn:Asymhyp}
\del_{t}\begin{pmatrix} A_{0} \\ A_{1} \end{pmatrix} + \begin{pmatrix} 0 & -e^{\alpha} \\ -e^{\alpha} & 0 \end{pmatrix} \del_{\theta}\begin{pmatrix} A_{0} \\ A_{1} \end{pmatrix}  -\alpha_{0}\begin{pmatrix} A_{0} \\ A_{1} \end{pmatrix} =& \begin{pmatrix} -A_{0} -4A_{0}U_{0} +4A_{1}U_{1} \\ 0 \end{pmatrix}, \\
\label{eqn:Usymhyp}
\del_{t}\begin{pmatrix} U_{0} \\ U_{1} \end{pmatrix} + \begin{pmatrix} 0 & -e^{\alpha} \\ -e^{\alpha} & 0 \end{pmatrix} \del_{\theta}\begin{pmatrix} U_{0} \\ U_{1} \end{pmatrix}  -\alpha_{0}\begin{pmatrix} U_{0} \\ U_{1} \end{pmatrix} =& \begin{pmatrix} \frac{1}{2} + \frac{1}{2}e^{4U+2t}(A_{0}^{2} - A_{1}^{2}) +U_{0} + \frac{1}{2}\alpha_{0}\\ 0 \end{pmatrix}.
\end{align}

\subsection{A First Order Formulation of the Scalar Field Equation}
The equation of motion  \eqref{eqn:Tij_divergence}
for the scalar field is equivalent to the wave equation $g^{ab}\nabla_{a}\nabla_{b}\phi = 0$ which, using the Gowdy metric \eqref{eqn:gowdymetricA}, 
can be expressed as
\begin{align*}
\del_{tt}\phi = \big(e^{2\alpha}\del_{\theta}\alpha\del_{\theta}\phi+e^{2\alpha}\del_{\theta\theta}\phi + \del_{t}\phi + \del_{t}\phi\del_{t}\alpha \big).
\end{align*}
Introducing the variables \begin{align}
\label{eqn:firstordervariables_phi}
\phi_{0} = \del_{t}\phi, \quad \phi_{1} = e^{\alpha}\del_{\theta}\phi,
\end{align}
allows us to write the scalar field equation 
in the first order form as
\begin{align}
\label{eqn:phisymhyp}
\del_{t}\begin{pmatrix} \phi_{0} \\ \phi_{1} \end{pmatrix} + \begin{pmatrix} 0 & -e^{\alpha} \\ -e^{\alpha} & 0 \end{pmatrix} \del_{\theta}\begin{pmatrix} \phi_{0} \\ \phi_{1} \end{pmatrix}  -\alpha_{0}\begin{pmatrix} \phi_{0} \\ \phi_{1} \end{pmatrix} =& \begin{pmatrix} \phi_{0} \\ 0\end{pmatrix}.
\end{align}

\subsection{Euler Equations}
\label{sec:EulerEquations}
Following \cite{BMO:2023}, the Euler equations \eqref{eqn:Fluid_Tij_divergence}  can 
be expressed in Gowdy symmetry as 
\begin{align} \label{Euler-B}
B^{0}\del_{0}V + B^{1}\del_{1}V = F, 
\end{align}
where
\begin{align*}
V =& \begin{pmatrix}\rho \\ v_{1} \end{pmatrix},\\
B^{0} =& \begin{pmatrix} \frac{K}{\rho+K\rho}\left(g_{11}+(v_{1})^{2}\right) & Kv_{1} \\ Kv_{1} & \rho+K\rho\end{pmatrix}, \\
B^{1} =& (-v_{0})\begin{pmatrix} \frac{K}{\rho+K\rho}v_{1} & K \\ K & (\rho+K\rho)\frac{v_{1}}{g_{11}+(v_{1})^{2}}\end{pmatrix},
\end{align*}
and
\begin{align*}
&F = \frac{1}{2}(-v_{0})\begin{pmatrix} K\left(2g^{11}\del_{1}g_{11}-g^{ab}\del_{1}g_{ab}\right)v_{1} \\(\rho+K\rho)\left(\frac{(v_{1})^{2}}{g_{11}+(v_{1})^{2}}g^{11}\del_{1}g_{11} - g^{00}\del_{1}g_{00}\right)\end{pmatrix} +\frac{K}{2} \begin{pmatrix} (v_{1})^{2}g^{11}\del_{0}g_{11}-\big(g_{11}+(v_{1})^{2}\big)g^{IK}\del_{0}g_{IK} \\ 0 \end{pmatrix}.
\end{align*}
To facilitate the study of the fluid near the big bang singularity, which is now located at $t=\infty$, we remove the leading order behaviour in $t$ from the fluid density $\rho$ by employing a renormalised density $\rhot$ defined by 
\begin{align}
\label{eqn:rhotdef}
\rho &= e^{\frac{3(1+K)}{2}t}\rhot.
\end{align}

Next, we differentiate $(\rho,v_1)$ to obtain the identities
\begin{align*}
\del_{t}\begin{pmatrix}\rho \\ v_{1} \end{pmatrix}= P\del_{t}\begin{pmatrix}\rhot \\ v_{1}\end{pmatrix} +Z\AND
\del_{\theta}\begin{pmatrix}\rho \\ v_{1} \end{pmatrix}= P\del_{\theta}\begin{pmatrix}\rhot \\ v_{1}\end{pmatrix},
\end{align*}
where 
\begin{equation*}
P = \begin{pmatrix} e^{\frac{3(1+K)}{2}t} & 0 \\ 0 & 1\end{pmatrix} \AND
Z = \begin{pmatrix} \frac{3(1+K)}{2} e^{\frac{3(1+K)}{2}t}\rhot  \\ 0 \end{pmatrix}.
\end{equation*}
Using these identities, we can express the Euler equations \eqref{Euler-B} as
\begin{align}
\label{eqn:Eulertilde}
\tilde{B}^{0}\del_{t}\mathbf{\tilde{V}} + \tilde{B}^{1}\del_{\theta}\mathbf{\tilde{V}} = F_{\mathbf{\Vt}},
\end{align}
where
\begin{equation*}
\mathbf{\tilde{V}} = (\rhot,v_1)^T
\end{equation*}
and
$\tilde{B}^{0}:= P^{\text{T}}B^{0}P$, $\tilde{B}^{1}:=P^{\text{T}}B^{0}P$ and $F_{\mathbf{\Vt}}:=P^{\text{T}}(F - B^{0}Z)$ take the form
\begin{align*}
\tilde{B}^{0} & = \begin{pmatrix} \frac{ K (e^{2 \eta} + e^{2 U} (v_{1})^2)}{e^{2 U} (1 + K) \rhot} &  K v_{1}\\  K v_{1} &  (1 + K) \rhot \end{pmatrix} ,\\ 
\tilde{B}^{1} &= \begin{pmatrix}
- \frac{K v_{1} (e^{2 \alpha } (e^{-2 U + 2 \eta } + (v_{1})^2))^{1/2}}{(1 + K) \rhot} & -  K (e^{2 \alpha } (e^{-2 U + 2 \eta } + (v_{1})^2))^{1/2}\\
-  K (e^{2 \alpha } (e^{-2 U + 2 \eta } + (v_{1})^2))^{1/2} & - \frac{e^{ 2 \alpha } (1 + K) \rhot v_{1}}{(e^{2 \alpha } (e^{-2 U + 2 \eta } + (v_{1})^2))^{1/2}}
\end{pmatrix}
\intertext{and}
F_{\mathbf{\Vt}} &= \begin{pmatrix}
- \frac{ K (e^{2 U} (v_{1})^2 - 2 e^{2 U} v_{1} (e^{2 \alpha } (e^{-2 U + 2 \eta } + (v_{1})^2))^{1/2} \del_{\theta}\alpha + e^{2 \eta } (1 - 2 \del_t U+ 2 \del_{t}\eta))}{2 e^{2 U}}\\
\tfrac{1}{2}  (1 + K) \rhot (-3 K v_{1} + \frac{2 (e^{2 \alpha } (e^{-2 U + 2 \eta } + (v_{1})^2))^{1/2} (e^{2 U} (v_{1})^2 \del_{\theta}\alpha + e^{2 \eta } (- \del_{\theta}U + \del_{\theta}\alpha+ \del_{\theta}\eta))}{e^{2 \eta } + e^{2 U} (v_{1}{})^2})
\end{pmatrix},
\end{align*}
respectively.

\subsection{The Complete Evolution System}
Combining \eqref{eqn:alphaevo_1}, \eqref{eqn:Hamiltonian_constraint1}, \eqref{eqn:Asymhyp}, \eqref{eqn:Usymhyp}, \eqref{eqn:phisymhyp}, and \eqref{eqn:Eulertilde} gives the system of equations which we solve numerically. These equations can be expressed in matrix form as
\begin{align}
\label{eqn:EinsteinEuler1}
\begin{pmatrix} \mathbb{I} & 0 & 0 & 0\\ 0 & \mathbb{I} & 0 & 0 \\ 0 & 0 & \tilde{B}^{0} &0 \\ 0 & 0 & 0 & \mathring{B}^{0} \end{pmatrix} \del_{t} \begin{pmatrix} \textbf{A} \\ \textbf{U} \\ \tilde{\textbf{V}} \\ \boldsymbol{\phi} \end{pmatrix} +\begin{pmatrix} \bar{B}^{1} & 0 & 0 & 0\\ 0 & \bar{B}^{1} & 0 & 0 \\ 0 & 0 & \tilde{B}^{1} &0 \\ 0 & 0 & 0 & \mathring{B}^{1} \end{pmatrix} \del_{\theta} \begin{pmatrix} \textbf{A} \\ \textbf{U} \\ \tilde{\textbf{V}} \\ \boldsymbol{\phi} \end{pmatrix} &= \begin{pmatrix} \alpha_{0} & 0 & 0 & 0\\ 0 & \alpha_{0}& 0 & 0 \\ 0 & 0 & 0 &0 \\ 0 & 0 & 0 & \alpha_{0} \end{pmatrix}\begin{pmatrix} \textbf{A} \\ \textbf{U} \\ \tilde{\textbf{V}} \\ \boldsymbol{\phi} \end{pmatrix} + \begin{pmatrix} F_{\mathbf{A}} \\ F_{\mathbf{U}} \\ F_{\mathbf{\tilde{V}} } \\ F_{\boldsymbol{\phi}}\end{pmatrix}, \\
\label{eqn:EinsteinEuler2}
\del_{t} \begin{pmatrix} \alpha \\ \eta \\ A \\ U \\ \phi \end{pmatrix} &= \begin{pmatrix} F_{\alpha} \\ F_{\eta} \\ A_{0} \\ U_{0} \\ \phi_{0} \end{pmatrix}, 
\end{align}
where
\begin{align*}
\mathbf{A} =& (A_{0}, A_{1})^T, \;\;\mathbf{U} = (U_{0},U_{1})^T, \;\; \boldsymbol{\phi} = (\phi_{0}, \phi_{1})^T, \\ 
\bar{B}^{1} =& \begin{pmatrix} 0 & -e^{\alpha} \\ -e^{\alpha} & 0 \end{pmatrix}, \;\; \mathring{B}^{0} = \begin{pmatrix} 1 & 0 \\ 0& 1 \end{pmatrix}, \;\; \mathring{B}^{1} = \begin{pmatrix} 0 & -e^{\alpha} \\ -e^{\alpha}  & 0 \end{pmatrix}, \\
F_{\mathbf{A}} =&\bigl( -A_{0} -4A_{0}U_{0} +4A_{1}U_{1}, 0 \bigr)^T, \;\; F_{\mathbf{U}} = \biggl( \frac{1}{2} + \frac{1}{2}e^{4U}(A_{0}^{2} - A_{1}^{2}) +U_{0} + \frac{1}{2}\alpha_{0}, 0 \biggr)^T,  \\
F_{\boldsymbol{\phi}} =&(\phi_{0}, 0)^T,  \;\; F_{\alpha} = -1 -e^{\frac{3(1+K)}{2}t -2U +2\alpha +2\eta}(K-1)\rhot,  \\
F_{\eta} =& -e^{\frac{3(1+K)}{2}t - 2U +2\alpha +2\eta}\rhot - \frac{1}{4}e^{2t+4U}(A_{1}^{2} + A_{0}^{2}) \nonumber \\
&+ \frac{1}{2}\Big(-2e^{\frac{3(1+K)}{2}t+2\alpha}(1+K)\rhot v_{1}^{2} - 2U_{1}^{2} - \phi_{1}^{2}\Big) - U_{0}^{2}  - \frac{1}{2}\phi_{0}^{2},
\end{align*}
and $\mathbf{\Vt}$, $\Bt^0$, $\Bt^1$ and $F_{\mathbf{\Vt}}$ are as defined above in the preceding section. Furthermore, we note that the momentum constraint \eqref{eqn:Momentum_constraint1} takes the form
\begin{align} \label{eqn:Momentum_constraint2} 
\del_{\theta}\eta= - e^{3/2 (1 + K) t} (1 + K) \rhot v_{1} \sqrt{e^{2 \alpha } (e^{-2 U + 2 \eta } + (v_{1})^2)}  - \del_{\theta}\alpha - \tfrac{1}{2} e^{2 t + 4 U}A_{0}\del_{\theta}A   -2 U_{0}\del_{\theta}U - \phi_{0}\del_{\theta}\phi.
\end{align}

\section{FLRW Solutions}
\label{sec:FLRWsoln}
As discussed in the introduction, the main aim of this article is to study Gowdy symmetric perturbations of FLRW solutions (i.e.\ spatially homogeneous and isotropic) to the Einstein-Euler-scalar field equations. This requires us to first identify the FLRW solutions. To this end, we observe that a FLRW metric can be recovered from the Gowdy metric \eqref{eqn:gowdymetricA} by  setting $\eta = -t$, $U=-\frac{1}{2}t$, and $A=0$ and assuming that the remaining metric function $\alpha$ depends only on $t$. This gives a metric of the form 
\begin{align*}
g = e^{-t}(e^{2\alpha(t)}dt^{2} + d\theta^{2} + dy^{2} + dz^{2}).
\end{align*}
Clearly, this metric ansatz is both spatially homogeneous and isotropic, and  hence, if we can find a solution to the Einstein-Euler-scalar field equations of this form it must, by definition, be the FLRW solution. For matter variables $\rhot$, $v_{1}$ and $\phi$, spatial homogeneity and isotropy requires that $v_{1}=0$ and that $\rhot$ and $\phi$ depend only on $t$. For these choices, the Gowdy-symmetric Einstein-Euler scalar field equations \eqref{eqn:EinsteinEuler2} simplify to
\begin{align}
\label{eqn:homog1}
\del_{t}\rhot &= 0, \\
\label{eqn:homog2}
\del_{t}\alpha + e^{\frac{1}{2}(3K+1)t + 2\alpha}(K-1)\rhot + 1 &= 0 , \\
\label{eqn:homog3}
2\phi_{0}^{2} +4e^{\frac{1}{2}(3K+1)t + 2\alpha}\rhot -3 &= 0, \\
\label{eqn:homog4}
2\phi_{0}^{2} + 4\del_{t}\alpha + 1 + 4e^{\frac{1}{2}(3K+1)t + 2\alpha}K\rhot +1&= 0, \\
\label{eqn:homog5}
\del_{t}\phi_{0} -\phi_{0}\del_{t}\alpha -\phi_{0} &= 0, \\
\label{eqn:homog6}
\del_{t}\phi &= \phi_{0}.
\end{align}

We begin solving the above system by first noting that \eqref{eqn:homog1} implies
\begin{align*}
\rhot = \frac{1}{\tilde{c}(1+K)}
\end{align*}
where  the constant $\tilde{c}>0$ can be freely chosen. 
Next, we integrate \eqref{eqn:homog2} to obtain
\begin{align*}
\alpha = -\frac{1}{2}\log\Big(\frac{4}{3}e^{\frac{3(K-1)}{2}t}\frac{1}{\tilde{c}(K+1)} -2c_{1}\Big) -t
\end{align*}
where $c_{1}$ is freely specifiable integration constant. 
Substituting this into \eqref{eqn:homog3} and setting $c_{1}=-1$, we solve for  $\phi_0=\del_{t}\phi$ and integrate the resulting expression while enforcing the initial condition\footnote{Only derivatives of $\phi$ appear in the field equations, hence there is no loss of generality from using this condition to choose our integration constant.} $\phi|_{t=0}=0$ to get
\begin{equation*}
\phi = \frac{2\sqrt{2}}{\sqrt{3}(1-K)}\biggl( \sinh^{-1}\biggl(\sqrt{\frac{3}{2}\tilde{c}(1+K)}e^{\frac{3(1-K)}{4}t}\biggr)-  \sinh^{-1}\biggl(\sqrt{\frac{3}{2}\tilde{c}(1+K)}\biggr) \biggr).
\end{equation*}
It is then straightforward to check that the above expressions for $\rhot$, $\alpha$, and $\phi$ also  satisfy the remaining equations \eqref{eqn:homog4}-\eqref{eqn:homog6}. 
From this, we conclude that for each choice of constant $\ct>0$, the following defines the FLRW solution of the Einstein-Euler-scalar field equations:
\begin{equation} \label{eqn:solnhomog}
\begin{aligned}
g &= e^{-t}\biggl(\frac{-3\tilde{c}e^{-2t}(1+K)}{4e^{\frac{3(K-1)}{2}t} + 6\tilde{c}(1+K)}\dd t^{2} +\dd\theta^{2}+\dd y^{2} + \dd z^{2}\biggr), \\ 
\rho &=  \frac{1}{\tilde{c}(1+K)}e^{\frac{3(1+K)}{2}t}, \\
v &= \sqrt{\frac{3\tilde{c}e^{-3t}(1+K)}{4e^{\frac{3(K-1)}{2}t} + 6\tilde{c}(1+K)}}\,dt, \\
\phi &= \frac{2\sqrt{2}}{\sqrt{3}(1-K)}\biggl( \sinh^{-1}\biggl(\sqrt{\frac{3}{2}\tilde{c}(1+K)}e^{\frac{3(1-K)}{4}t}\biggr)-  \sinh^{-1}\biggl(\sqrt{\frac{3}{2}\tilde{c}(1+K)}\biggr) \biggr).
\end{aligned} 
\end{equation}

\begin{rem}
    The Einstein-Euler-scalar field FLRW solution has no closed solution in terms of the standard FLRW coordinates\footnote{By standard FLRW coordinates, we mean metrics of the form $g= -dT^{2} + a(T)^{2}(dx^{2}+dy^{2}+dz^{2})$.} for arbitrary values of $K$ \cite{Faraoni:2021}. This can be seen by changing our areal time coordinate to the standard FLRW time, which  yields an expression in terms of a hypergeometric function,
\begin{align*}
    T &= \int \sqrt{\frac{3\tilde{c}e^{-3t}(1+K)}{4e^{\frac{3(K-1)}{2}t} + 6\tilde{c}(1+K)}}\, dt = -\frac{\sqrt{\frac{2}{3}} e^{-3 t} \, _2F_1\left(1,\frac{1}{2}+\frac{1}{1-K};\frac{2-K}{1-K};-\frac{2 e^{\frac{3}{2} (K-1) t}}{3 \tilde{c} (K+1)}\right)}{3 \sqrt{\frac{\tilde{c} (K+1) e^{-3
   t}}{3 \tilde{c} (K+1)+2 e^{\frac{3}{2} (K-1) t}}}}.
\end{align*}
In particular, 
this expression can not be analytically inverted to obtain a closed solution $t=t(T)$.
\end{rem}

\section{Numerical Results Near FLRW}
\label{sec:NumericalResults}

\subsection{Numerical Setup}
The numerical method we employ to solve \eqref{eqn:EinsteinEuler1}-\eqref{eqn:EinsteinEuler2} in this article is closely related to the one we used in \cite{BMO:2023} to solve the Gowdy-symmetric Einstein-Euler equations in the expanding direction. Specifically, we use a $[0,2\pi]$ spatial computational spatial domain that is discretised with an equidistant grid consisting of $N$ grid points, and we employ periodic boundary conditions to enforce the $2\pi$-periodicity of the gravitational and matter fields. Spatial derivatives are discretised using $2^{\text{nd}}$ order central finite differences and time integration is performed using a standard $4^{\text{th}}$ order Runge-Kutta method. As a consequence, our code is second order accurate.  

\subsubsection{Initial Data}
Since we are using the Hamiltonian constraint \eqref{eqn:Hamiltonian_constraint1} to evolve $\eta$, the constraints that our initial data for the system \eqref{eqn:EinsteinEuler1}-\eqref{eqn:EinsteinEuler1} must satisfy consist of the momentum constraint
\eqref{eqn:Momentum_constraint2}  and the constraints \eqref{eqn:firstordervariablesA_U} and  \eqref{eqn:firstordervariables_phi} that arise from the definition of the first order variables $A_{1}$, $U_{1}$ and $\phi_{1}$. The choice of initial data \eqref{eqn:numericalID} below ensures all these constraints are satisfied initially at $t=0$. 
Additionally, we choose the fluid's initial spatial velocity $v_1$ so that it vanishes at least one point on the initial hypersurface at $t=0$. This is necessary to generate the tilt-instability that leads to the formation of spikes in the fractional density contrast $\frac{\del_\theta\rho}{\rho}$ and  ultimately blow-up on the big bang singularity at $t=\infty$. Following \cite{BMO:2023,MarshallOliynyk:2022}, we ensure that the spatial fluid velocity $v_1$ vanishes at $t=0$ by setting it equal to a sinusoidal function with a small amplitude parameter $a$. For the remainder of this section, we employ initial data of the form\footnote{Here we have set the constant $\tilde{c}=1$.} 
\begin{equation}
\begin{aligned}
\label{eqn:numericalID}
\mathring{\alpha} &= -\log\biggl(-a\cos(\theta)+\sqrt{\frac{4}{3(K+1)}+2}\biggr), \;\; a < \sqrt{\frac{4}{3(K+1)}+2}, \\
\mathring{v_{1}} &= a\sin(\theta), \\
\mathring{\rhot} &= \frac{1}{(K+1)\sqrt{e^{2(\eta-U)}+v_{1}^{2}}}, \\
\mathring{\eta} &= -(2f-1)c\sin(\theta) - d\biggl(\frac{3\sqrt{1+K}}{\sqrt{4+6(1+K)}}+b\biggr)\sin(\theta), \\
\mathring{U} &= c\sin(\theta), \\
\mathring{U}_{0} &= -\frac{1}{2} + f, \\
\mathring{\phi}_{0} &= \frac{3\sqrt{1+K}}{\sqrt{4+6(1+K)}}+b, \\
\mathring{\phi} &= d\sin(\theta), \\
\mathring{A}_{0} &= 0, \\
\mathring{A} &= k\sin(\theta) +b, \\
\mathring{A}_{1} &= e^{\alpha}\del_{\theta}A, \\
\mathring{U}_{1} &= e^{\alpha}\del_{\theta}U, \\
\mathring{\phi}_{1} &= e^{\alpha}\del_{\theta}\phi.
\end{aligned}
\end{equation}
where $a$, $b$, $c$, $d$, $f$, and $k$ are constants to be specified. Initial data of this form can be considered as a perturbation of FLRW initial data
provided that the constants $a$, $b$, $c$, $d$, $f$, and $k$ are chosen sufficiently close to zero. This follows from the fact that setting $a=b=c=d=f=k=0$ in \eqref{eqn:numericalID} produces homogeneous and isotropic (i.e.\ FLRW) initial data. If the size of the parameters $a,b,c,d,f$, and $k$ are too large the system is found to become unstable almost immediately. That is, within a small amount of timesteps the variables develop steep gradients and produce numerical errors.  Throughout this section, we focus exclusively on initial data with small amplitudes. In particular, all the plots in this section have been generated with $a=b=c=d=f=k=0.01$, with the exception of Section \ref{sec:codevalidation} where we set $a=b=c=d=f=k=0$.

\subsubsection{Code Tests}
\begin{figure}[htbp]
\centering
\subfigure[Subfigure 1 list of figures text][$v_{1}$]{
\includegraphics[width=0.4\textwidth]{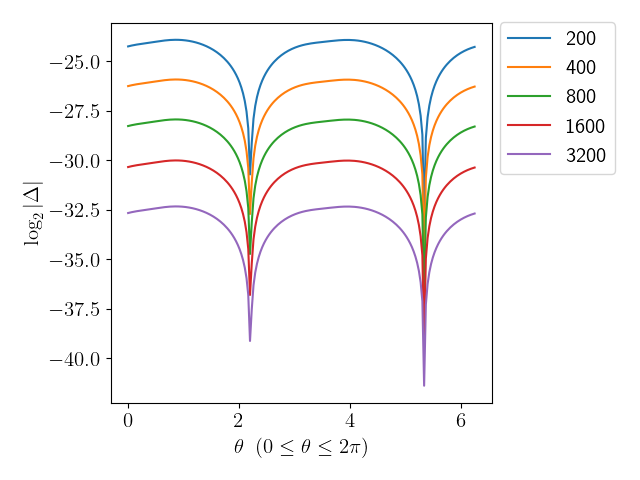}
\label{fig:v_convergence}}
\subfigure[Subfigure 2 list of figures text][$\phi$]{
\includegraphics[width=0.4\textwidth]{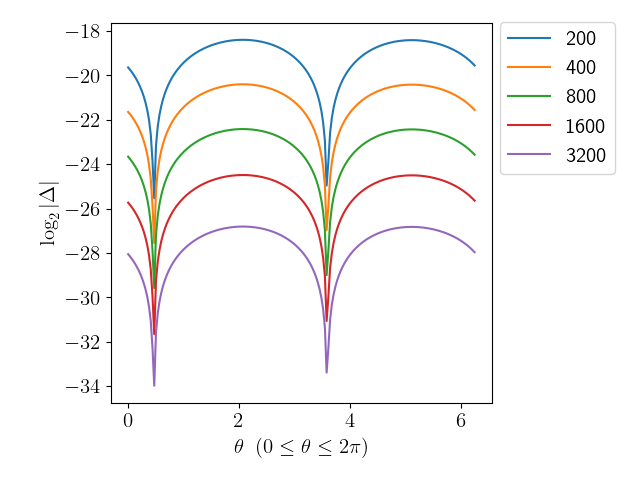}
\label{fig:phi_convergence}}
\caption{Convergence plots of $v_{1}$ and $\phi$ at $t = 15.07$,  $K=0.1$.}
\end{figure}

The second order accuracy of our code has been verified with convergence tests involving perturbations of FLRW solutions using resolutions of $N =$ $200$, $400$, $800$, $1600$, $3200$, and $6400$ grid points. Following \cite{BMO:2023}, we have estimated the numerical discretisation error $\Delta$ by taking the $\log_{2}$ of the absolute value of the difference between each simulation and the highest resolution run. The results for $v_{1}$ and $\phi$  are shown\footnote{We have performed convergence tests for all other variables and confirmed second order convergence. These plots are omitted here for brevity.} in Figures \ref{fig:v_convergence}-\ref{fig:phi_convergence} from which the second order convergence is clear. \newline

\begin{figure}[htbp]
\centering
\subfigure[Subfigure 1 list of figures text][$\log_{2}(\|C_{1}\|_{2}+\|C_{2}\|_{2}+\|C_{3}\|_{2}+\|C_{4}\|_{2})$ Constraint]{
\includegraphics[width=0.4\textwidth]{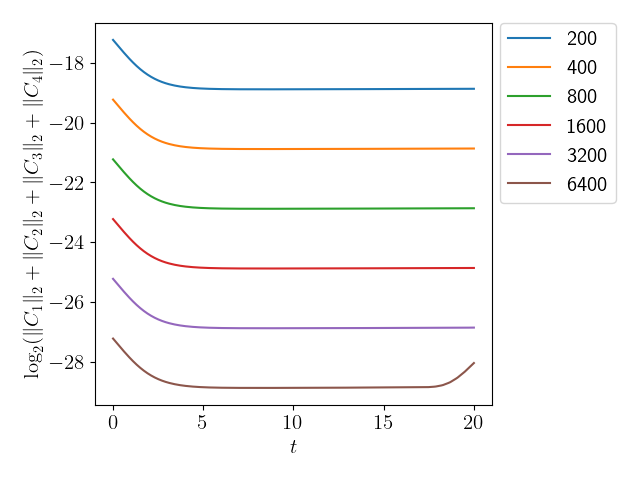}
\label{fig:subfigCsum}}
\subfigure[Subfigure 2 list of figures text][$\log_{2}(\|C_{5}\|_{2})$ Constraint]{
\includegraphics[width=0.4\textwidth]{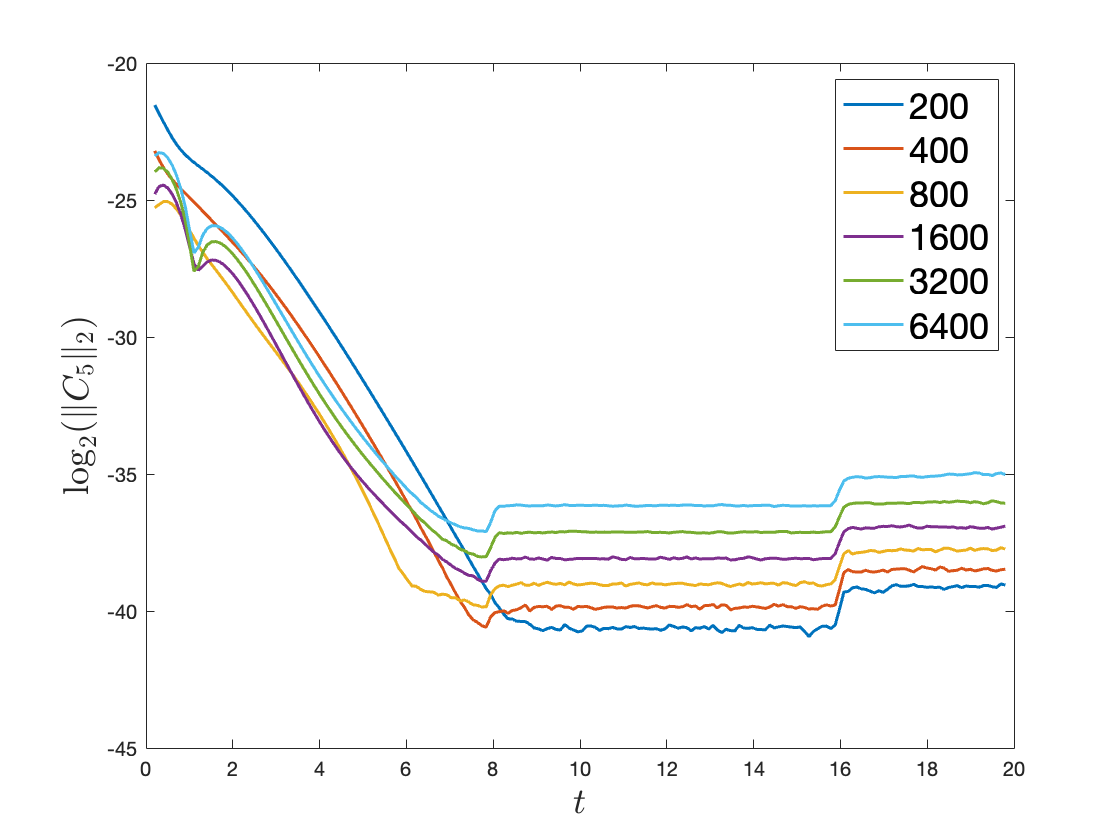}
\label{fig:subfigC5}}
\caption{Convergence plots of the constraint quantities, $K=0.1$. The system was evolved until $t=20$. }
\end{figure}

We can use a similar procedure to measure the level of constraint violation during the evolution of the system. Beginning with the momentum constraint \eqref{eqn:Momentum_constraint2}, we define the quantity
\begin{align}
\label{eqn:momentumconstraintquantity}
    C_{1} = -\del_{\theta}\eta - e^{3/2 (1 + K) t} (1 + K) \rhot v_{1} \sqrt{e^{2 \alpha } (e^{-2 U + 2 \eta } + (v_{1})^2)} 
- \del_{\theta}\alpha - \tfrac{1}{2} e^{2 t + 4 U}A_{0}\del_{\theta}A   -2 U_{0}\del_{\theta}U - \phi_{0}\del_{\theta}\phi.
\end{align}
Clearly, $C_{1}=0$ means that the momentum constraint is identically satisfied. The quantity $\log_2 \norm{C_1}_{2}$ can therefore be understood as the violation error of the momentum constraint as a function of time.  In a similar manner, we can also define constraint violation quantities from the definitions of our first order variables $A_{1}$ and $U_{1}$, and from the wave equation \eqref{eqn:Etawave} for $\eta$ as follows
\begin{align*}
    C_{2} &= A_{1}-e^{\alpha}\del_{\theta}A, \\
    C_{3} &= U_{1}-e^{\alpha}\del_{\theta}U, \\
    C_{4} &= \phi_{1} - e^{\alpha}\del_{\theta}\phi, \\
    C_{5} &= -\del_{tt}\eta -\frac{1}{4}e^{-2U}\Big(4e^{2\alpha+2\eta + \frac{3(1+K)}{2}t}K\rhot + e^{2t+6U+2\alpha}(\del_{\theta}A)^{2} - 4e^{2U+2\alpha}(\del_{\theta}U)^{2} - 4e^{2U+2\alpha}(\del_{\theta}\alpha)^{2} \nonumber \\
&\quad - 4e^{2U+2\alpha}\del_{\theta}\alpha\del_{\theta}\eta - 2e^{2U+2\alpha}(\del_{\theta}\phi)^{2} - 4e^{2U +2\alpha}\del_{\theta\theta}\alpha \nonumber \\
&\quad - 4e^{2U}(\del_{t}U)^{2} - 4e^{2U}\del_{t}\alpha\del_{t}\eta + 2e^{2U}(\del_{t}\phi)^{2}\Big)
\end{align*}
The second time derivative of $\eta$ for $C_{5}$ is calculated numerically using a fourth order finite difference stencil for the second derivative
\begin{align}
\label{eqn:time_finitediff}
(\del_{tt}\eta)_{i,j} = \frac{-\eta_{i-2,j}+16\eta_{i-1,j}-30\eta_{i,j}+16\eta_{i+1,j}-\eta_{i+2,j}}{12(\Delta t)^{2}},
\end{align}
where $\eta_{i,j}$ denotes the value of $\eta$ at the i\textsuperscript{th} timestep and j\textsuperscript{th} spatial grid point and $\Delta t$ is the timestep size, while the first time derivatives of $\alpha$ and $\eta$ in $C_{5}$ are calculated using their evolution equations \eqref{eqn:alphaevo_1} and \eqref{eqn:Hamiltonian_constraint1} respectively.
We observe the expected second order convergence for the quantities $\log_{2}(\|C_{1}\|_{2}+\|C_{2}\|_{2}+\|C_{3}\|_{2}+\|C_{4}\|_{2})$, shown in Figure \ref{fig:subfigCsum}. It should be noted that we have been unable to achieve convergence for the constraint quantity $\log_{2}(\|C_{5}\|_{2})$, plotted in Figure \ref{fig:subfigC5}. Although this constraint does not converge, the overall constraint violation becomes small and approaches the limit of numerical accuracy for a scheme using 2nd order finite difference stencil (approximately $10^{-13}$).  Even though the constraints are satisfied at the initial time by virtue of our choice of initial data \eqref{eqn:numericalID} so that $C_1=C_2=C_3=C_4=0$, we note the numerical values are not exactly zero, even at the initial time $t=0$, as the derivatives in $C_1$, $C_{2}$, $C_{3}$, and $C_{4}$ are approximated by finite differences. It should also be noted that, due to our use of the stencil \eqref{eqn:time_finitediff}, the first and last two timesteps in calculating $C_5$ have been removed from Figure \ref{fig:subfigC5}. 

\begin{figure}
    \centering
    \includegraphics[width=0.4\textwidth]{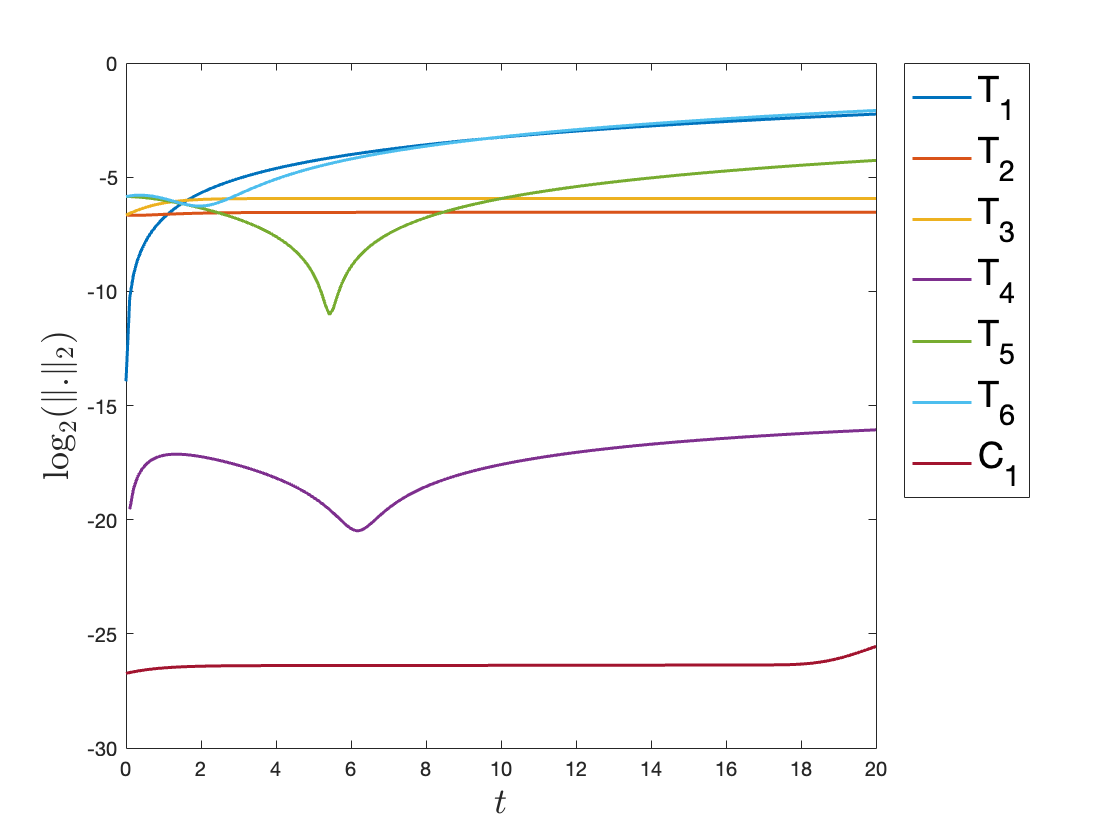}
    \caption{Comparison of the $L_{2}$ norm of the individual terms in momentum constraint and the combined constraint quantity $C_{1}$. $N=6400$, $K=0.1$.}
    \label{fig:constraint_termcompare}
\end{figure}

As a further check on the accuracy of the code, we have also compared the size of each individual term in a constraint with the total constraint violation. From this we can conclude that the actual constraint violation is small (as opposed to each individual term being small). To this end we consider $C_1$ and separate it into six terms as follows:
\begin{align}
\label{eqn:T1-constraint}
    T_{1} &= -\del_{\theta}\eta, \\
    T_{2} &= -e^{3/2 (1 + K) t} (1 + K) \rhot v_{1} \sqrt{e^{2 \alpha } (e^{-2 U + 2 \eta } + (v_{1})^2)}, \\
    T_{3} &= - \del_{\theta}\alpha, \\
    T_{4} &= - \tfrac{1}{2} e^{2 t + 4 U}A_{0}\del_{\theta}A, \\
    T_{5} &= -2 U_{0}\del_{\theta}U, \\
    \label{eqn:T6-constraint}
    T_{6} &= - \phi_{0}\del_{\theta}\phi
\end{align}
For the constraint violations $C_1$ to be actually small, we expect that the norm of each individual term \eqref{eqn:T1-constraint}-\eqref{eqn:T6-constraint} should be larger than the norm of the total constraint violation $C_1$ since this indicates that a cancellation among the terms in the sum is occurring. Figure~\ref{fig:constraint_termcompare} demonstrates that this cancellation is happening for $C_1$. We observe similar behaviour for the other constraints, $C_{2}$, $C_{3},$ $C_{4}$, and $C_{5}$. From these observations, we conclude that the constraints are being preserved sufficiently well by our numerical scheme.

\subsubsection{Code Validation}\label{sec:codevalidation}
\begin{figure}
    \centering
    \includegraphics[width=0.4\textwidth]{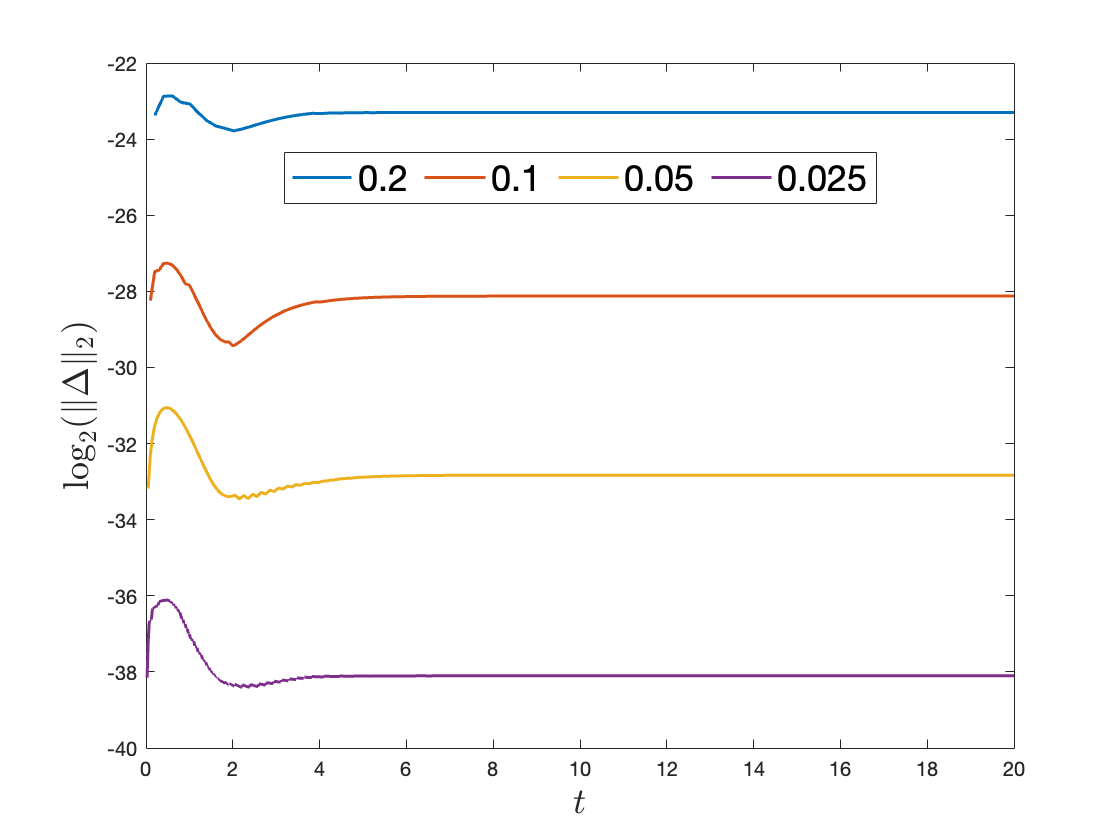}
    \caption{Convergence plot of the $L_{2}$ norm of $\phi_{0}-\phi_{0}^{\text{exact}}$ for several different values of the timestep $\Delta t$. All evolutions used $N=200$ and $K=0.2$. The system was evolved until $t=20$.}
    \label{fig:phiexact}
\end{figure}

A simple way to test the validity of our code is to compare our numerical solution with the exact FLRW solution \eqref{eqn:solnhomog}. For this convergence test, we employ the following initial data, which is obtained by setting $a=b=c=d=f=k=0$,
\begin{equation*}
\begin{aligned}
\rhot &= \frac{1}{K+1}, \\
\alpha &= -\log\Big(\sqrt{\frac{4}{3(K+1)}+2}\Big), \\
\phi_{0} &= \frac{3\sqrt{1+K}}{\sqrt{4+6(1+K)}},\\
U_{0} &= -\frac{1}{2}, \\
A&=A_{1}=A_{0}=U=U_{1}=\eta=\phi=\phi_{1}=v_{1}=0.
\end{aligned}
\end{equation*}
Due to the homogeneity of the solution, the order of convergence only depends on our time stepping method, which is fourth order accurate. Our scheme displays the expected convergence rate, shown for $\phi_{0}$ in Figure \ref{fig:phiexact}.

\subsection{Numerical Behaviour}
\label{sec:numericalbehaviour}

\subsubsection{Asymptotic Behaviour and Approximations}
\label{sec:asympbehav}
Before we present our numerical results, we first derive the expected asymptotics for the solutions from the evolution equations through a heuristic analysis. In particular, we will justify  \eqref{eq:specialimplsolfluid.Intro}; recall that $\bar t$ and $t$ are related by \eqref{eq:timetransformation} and that the big bang asymptotics corresponds to the limits $\bar t\searrow 0$ and $t\rightarrow\infty$, respectively. To this end, we suppose that, near the singularity, the Einstein-Euler-scalar field system is well-approximated by an FLRW solution of Einstein-scalar field equations. It follows from \eqref{eqn:solnhomog} that this metric takes the asymptotic form
\begin{equation}
    \label{eq:asymptoticFLRWmetric}
    g = e^{-t}\biggl(-\frac{1}{2}e^{-2t}\dd t^{2} +\dd\theta^{2}+\dd y^{2} + \dd z^{2}\biggr),
\end{equation}
which corresponds to the metric variables
\begin{equation*}
    U=-t/2,\quad \eta=-t,\quad \alpha=-t-\log\sqrt{2},\quad A=0.
\end{equation*}
\begin{rem}
 Employing the time coordinate $T=\frac{\sqrt{2}}{3}e^{-\frac{3}{2}t}$ allows us to express the metric   \eqref{eq:asymptoticFLRWmetric} in the standard FLRW form
 \begin{align*}
    g = -d T^2+\frac{3^{\frac{2}{3}}}{2^{\frac{1}{3}}}T^{\frac{2}{3}}\bigl(d\theta^2+dy^2+dz^2\bigr).
\end{align*}
As expected for a scalar field solution, the scale factor is proportional to $T^{\frac{1}{3}}$.
\end{rem}
Furthermore, we assume that the fluid part of the solution is asymptotically governed by the Euler equations on the background \eqref{eq:asymptoticFLRWmetric} with negligible spatial derivative terms. For each spatial point, we therefore have that
\begin{equation*}
    \tilde{B}^{0}  = \begin{pmatrix} \frac{ K (e^{-t} +  (v_{1})^2)}{(1 + K) \rhot} & K v_{1}\\ K v_{1} & (1 + K) \rhot \end{pmatrix} ,\quad \tilde{B}^{1}\del_\theta \mathbf{\Vt}  = 0,\quad
    F_{\mathbf{\Vt}} = \begin{pmatrix}
    - \frac{ K}2  (v_{1})^2   \\
    -\tfrac{3}{2}  (1 + K)K \rhot v_{1} 
    \end{pmatrix},
\end{equation*}
and thus,  the Euler equations \eqref{eqn:Eulertilde} reduce to
\begin{equation}
    \label{eq:asymptfluideqs}
    \partial_t \mathbf{\Vt}
    =  -\frac 12\frac 1{e^{-t} +  (1-K)(v_{1})^2}
    \begin{pmatrix}
          (1 + K)(1   -3K)(v_{1})^2\rhot  \\
        K\Bigl(  2(v_{1})^2    +{3} e^{-t}\Bigr)  v_{1} 
    \end{pmatrix}.
\end{equation}
It is then straightforward to show that for each spatial point the second equation in \eqref{eq:asymptfluideqs} implies the following implicit solution, cf.\ \eqref{eq:specialimplsolfluid.Intro},
\begin{equation}  
    \label{eq:specialimplsolfluid}
   \frac{e^{t/2}v_1(t)}{\bigl(1+e^t v_1^2(t)\bigr)^{K/2}}=\frac{\beta(t)}{\bigl(1-\beta^2(t)\bigr)^{(1-K)/2}} = c\, e^{-(3 K-1)  t/2}
\end{equation}
where $c\in\Rbb$ is an integration constant and
\begin{equation}
    \label{eq:defbeta}
    \beta(t)=\frac{e^{t/2}v_1(t)}{\sqrt{1+e^t v_1^2(t)}}.
\end{equation}
This solution is only valid for those $t\in\Rbb$ for which $|\beta(t)|<1$.
Recall from \eqref{v-Gowdy} and \eqref{eq:asymptoticFLRWmetric} that the fluid velocity tangent vector field (labelled by the same letter $v$ as the corresponding cotangent vector field in \eqref{eq:asymptoticFLRWmetric}) is given by
\begin{equation}
    v=-2e^{3t} v_0\partial_t+e^t v_1\partial_\theta=-\sqrt{2}e^{3t/2} v_0 e_0+e^{t/2} v_1 e_1
    =\frac 1{\sqrt{1-\beta^2(t)}}e_0+\frac {\beta(t)}{\sqrt{1-\beta^2(t)}}e_1,
\end{equation} 
where
\begin{equation}
    \label{eq:specialorthframe}
    e_0=-\sqrt{2}e^{3t/2}\partial_t,\quad e_1=e^{t/2}\partial_\theta,
\end{equation}
are frame vector fields orthonormal with respect to \eqref{eq:asymptoticFLRWmetric}.

The implicit solution \eqref{eq:specialimplsolfluid} can now be interpreted at each spatial point as follows. If $c=0$, we have $v_1(t)=\beta(t)=0$ because the right side of \eqref{eq:specialimplsolfluid} is identically zero \and we say the that fluid is \textit{orthogonal}. On the other hand, if $c\not=0$, we consider the three cases: $K\in (1/3,1]$, $K\in [0,1/3)$ and $K=1/3$. Now, if $c\not=0$ and $K\in (1/3,1]$,  $\beta(t)$ approaches zero as $t\rightarrow\infty$ because the right side of \eqref{eq:specialimplsolfluid} approaches zero in the limit $t\rightarrow\infty$. In this case, we refer to the fluid as \textit{asymptotically orthogonal}. Next, if $c\not=0$ and $K<1/3$, we have that $\beta^2(t)\rightarrow 1$ since the right side approaches infinity in the limit $t\rightarrow\infty$. Thus, the leading order behaviour of the fluid is a null vector and we call the fluid \textit{asymptotically extremely tilted}. Finally, if $K=1/3$, then $\beta(t)$ is a {non-zero} constant and hence the fluid has a non-vanishing asymptotically spatial velocity and we refer to the fluid as \textit{asymptotically tilted}.  As mentioned earlier, it is interesting to observe that the limit $t\rightarrow-\infty$ corresponds to switching the roles of the respective $K$ intervals: the fluid is asymptotically orthogonal if $K<1/3$ and asymptotically extremely tilted if $K\in (1/3,1]$. 

\begin{figure}[htbp]
\centering
\subfigure[Subfigure 1 list of figures text][$\rho$ at the 334th grid point (Blue) and tilted growth rate \eqref{eqn:TiltedDecay} (Orange)]{
\includegraphics[width=0.4\textwidth]{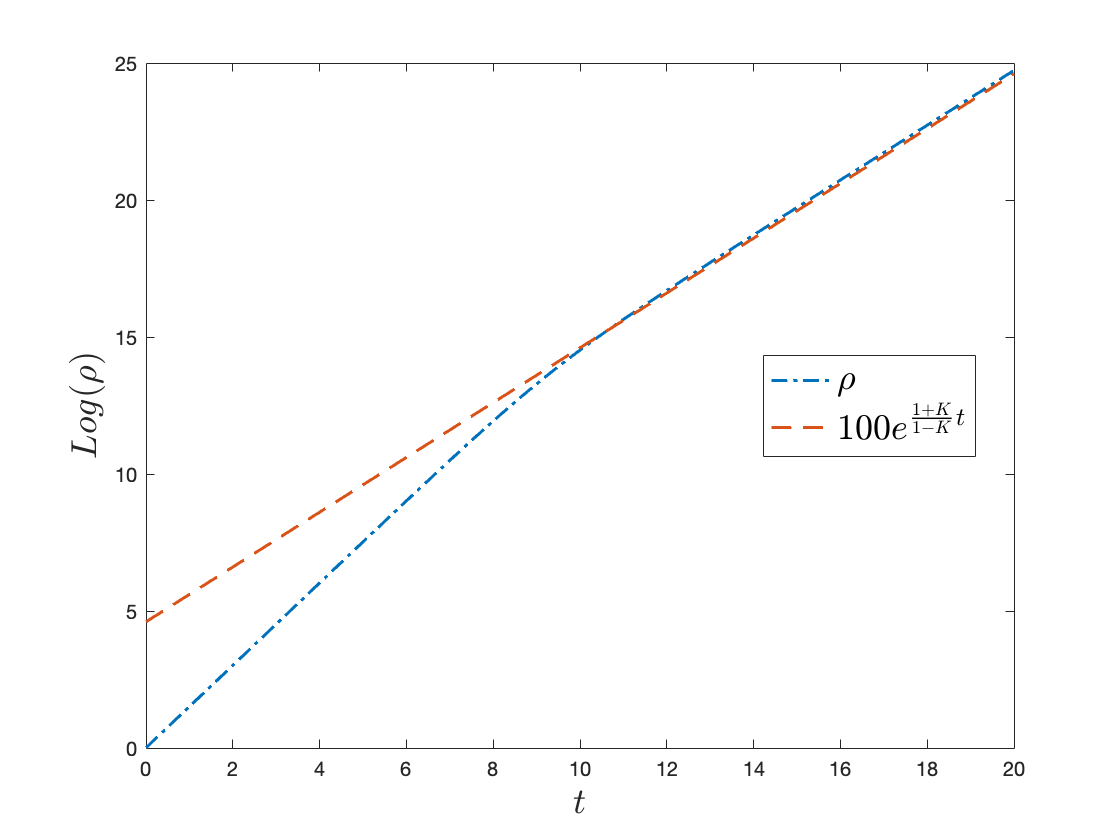}
\label{fig:334_asymptotics}}
\subfigure[Subfigure 2 list of figures text][$\rho$ at the 560th grid point (Blue) and orthogonal growth rate \eqref{eqn:IsotropicDecay} (Orange)]{
\includegraphics[width=0.4\textwidth]{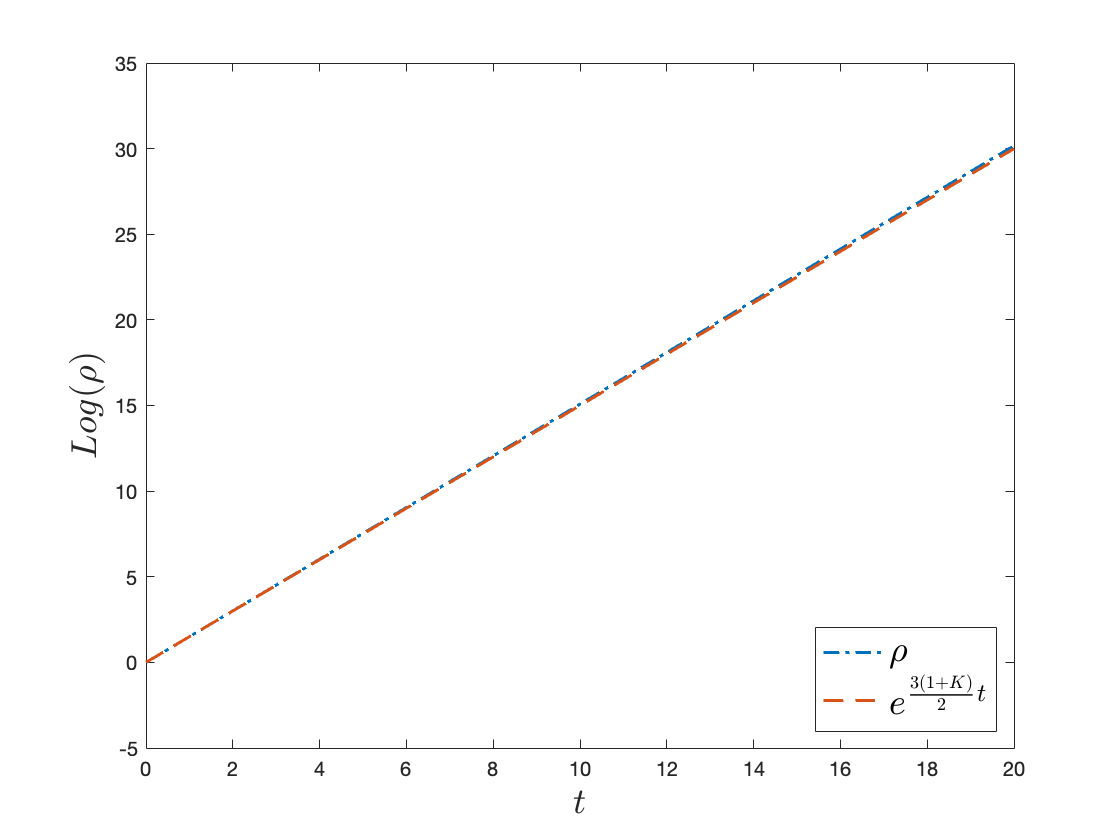}
\label{fig:560_asymptotics}}
\caption{Plot comparing the asymptotic behaviour of $\rho$ with asymptotically orthogonal and extremely tilted growth rates. $N=1000$, $K=0.1$.}
\label{fig:asymptotics}
\end{figure}

Using the first equation in \eqref{eq:asymptfluideqs} and \eqref{eqn:rhotdef} we conclude that
\begin{equation}
    \rho(t)=\rho_0\exp\Bigl(-\frac 12   (1 + K)(1   -3K)\int_{t_*}^t\frac {e^{s}v_{1}^2(s)}{1 +  (1-K)e^{s}v_{1}^2(s)}ds\Bigr)e^{\frac{3(1+K)}{2}t},
\end{equation}
where $\rho_0>0$ is an integration constant. All of this can be used now to derive the following asymptotics in the limit $t\rightarrow\infty$ from \eqref{eq:specialimplsolfluid}:
\begin{itemize}
    \item Orthogonal case ($c=0$):  
    \begin{equation}
    \label{eqn:IsotropicDecay}
        e^{t/2} v_1(t)=\beta(t)=0,\quad \rho(t)=O(e^{\frac{3(1+K)}{2}t}).
    \end{equation}
    \item Asymptotically orthogonal case ($c\not=0$, $K\in (1/3,1]$):
    \begin{equation}
        e^{t/2} v_1(t)=O(e^{-(3 K-1)  t/2}),\quad \beta(t)=O(e^{-(3 K-1)  t/2}),\quad \rho(t)=O(e^{\frac{3(1+K)}{2}t}).
    \end{equation}
    \item Asymptotically extremely tilted case ($c\not=0$, $K\in [0,1/3)$):
    \begin{equation}
    \label{eqn:TiltedDecay}
        e^{t/2} v_1(t)=O(e^{\frac{(1-3 K)  t}{2(1-K)}}),\quad 1-\beta^2(t)=O(e^{-\frac{1-3 K}{1-K}  t}),\quad \rho(t)=O(e^{\frac{1+K}{1-K}t}).
    \end{equation}
    \item Asymptotically tilted case ($c\not=0$, $K=1/3$):
    \begin{equation}
    \label{eqn:TiltedDecay.2}
        e^{t/2} v_1(t)=const,\quad \beta(t)=const, \quad \rho(t)=O(e^{2t}).
    \end{equation}
\end{itemize}
Our numerical scheme replicates the expected growth rates for all $K \in [0,1]$. In particular, for $K \in [0,\frac{1}{3})$ we observe the orthogonal growth rate \eqref{eqn:IsotropicDecay} near points where $v_{1}$ vanishes and the tilted rate \eqref{eqn:TiltedDecay} elsewhere, shown in Figure \ref{fig:asymptotics}. 

\subsubsection{Behaviour of the Ricci Scalar}
\begin{figure}
    \centering
    \includegraphics[width=0.4\textwidth]{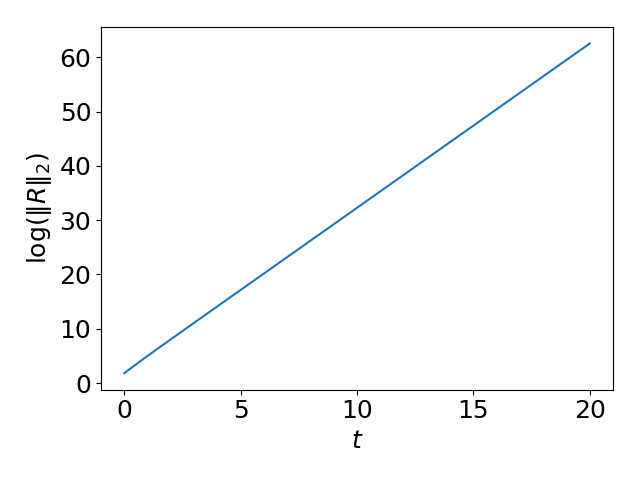}
    \caption{The natural logarithm of the $L_{2}$ norm of the Ricci scalar. $N=1000$, $K=0.1$.}
    \label{fig:ricci_scalar}
\end{figure}

We expect to see a curvature singularity in our numerical solutions as $t\rightarrow \infty$, which can be verified by the asymptotic behaviour of the Ricci scalar. Using the trace-reversed Einstein equation the Ricci scalar is given by 
\begin{align*}
    R &= -T = - e^{3/2 (1 + K) t} (-1 + 3 K) \rhot + \frac{e^{2 U} (\phi_{1}^{2} -\phi_{0}^{2})}{e^{2 \alpha  + 2 \eta }},
\end{align*}
where the $T$ is the trace of the stress-energy tensor.  As expected the Ricci curvature blows up as $t\rightarrow\infty$, shown in Figure \ref{fig:ricci_scalar}. 

\subsubsection{Behaviour of the density gradient} 
\label{sec:density_gradient}
\begin{figure}[htbp]
\centering
\subfigure[Subfigure 1 list of figures text][$t = 3.015$]{
\includegraphics[width=0.3\textwidth]{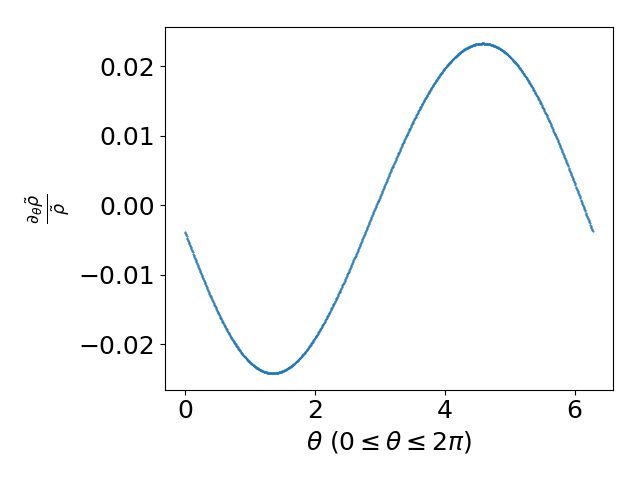}
\label{fig:subfigDC_t30}}
\subfigure[Subfigure 2 list of figures text][$t = 12.06$]{
\includegraphics[width=0.3\textwidth]{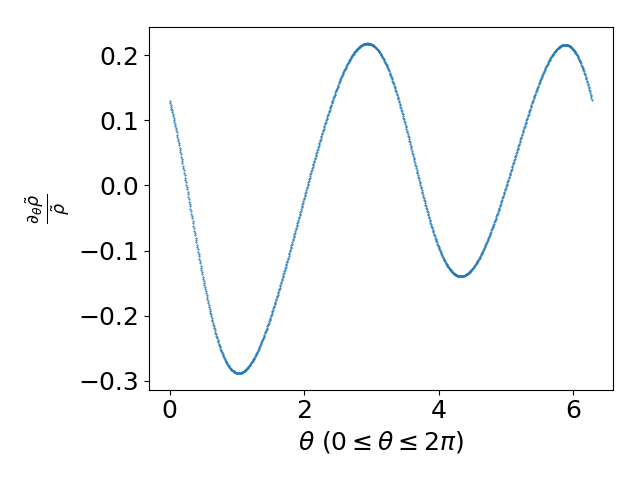}
\label{fig:subfigDC_t120}}
\subfigure[Subfigure 2 list of figures text][$t = 17.085$]{
\includegraphics[width=0.3\textwidth]{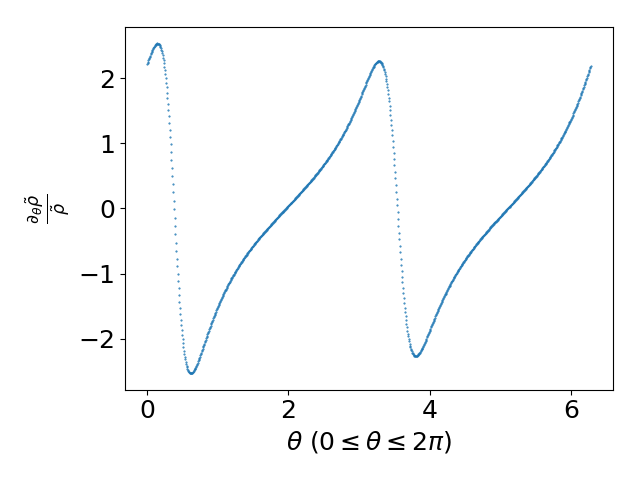}
\label{fig:subfigDC_t170}}
\caption{Density gradient $\frac{\del_{\theta}\rho}{\rho}$ at various times. $N=1000$, $K=0.1$.}
\label{fig:density_gradient}
\end{figure}

\begin{figure}[htbp]
\centering
\subfigure[Subfigure 1 list of figures text][$t = 5.025$]{
\includegraphics[width=0.3\textwidth]{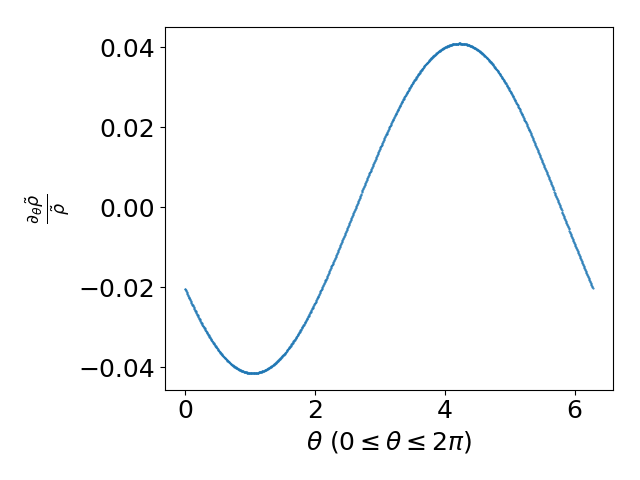}
\label{fig:subfigDC_t30_035}}
\subfigure[Subfigure 2 list of figures text][$t = 10.05$]{
\includegraphics[width=0.3\textwidth]{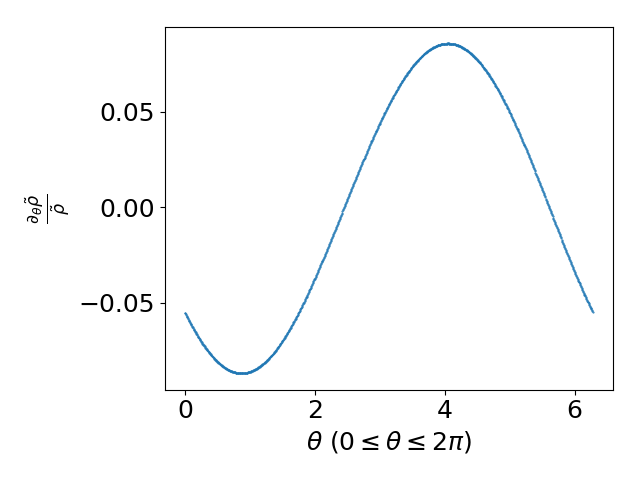}
\label{fig:subfigDC_t120_035}}
\subfigure[Subfigure 2 list of figures text][$t = 19.59$]{
\includegraphics[width=0.3\textwidth]{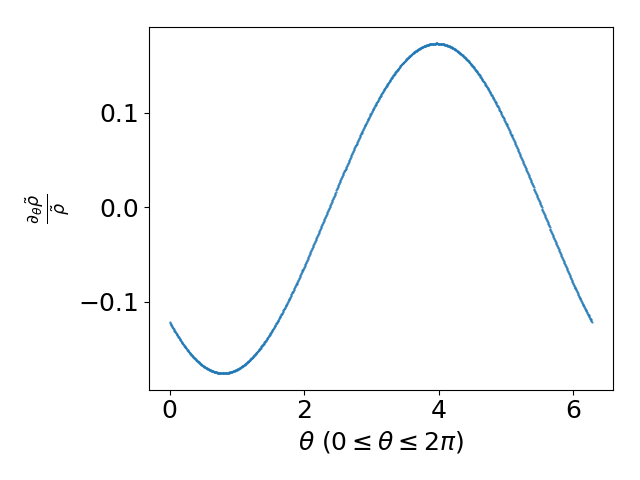}
\label{fig:subfigDC_t170_035}}
\caption{Density gradient $\frac{\del_{\theta}\rho}{\rho}$ at various times. $N=1000$, $K=0.35$.}
\label{fig:density_gradient_035}
\end{figure}

\begin{figure}[htbp]
\centering
\subfigure[Subfigure 1 list of figures text][$t = 3.015$]{
\includegraphics[width=0.3\textwidth]{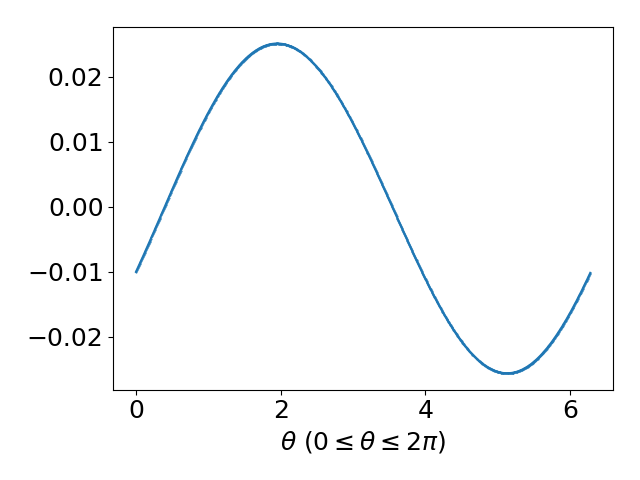}
\label{fig:TiltStep_t30}}
\subfigure[Subfigure 2 list of figures text][$t = 13.06$]{
\includegraphics[width=0.3\textwidth]{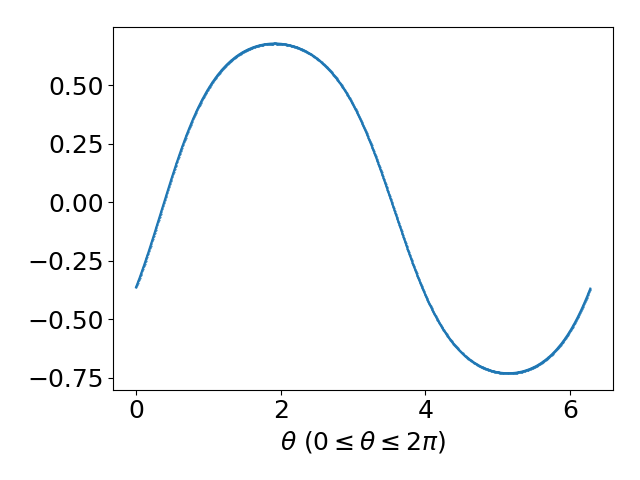}
\label{fig:TiltStep_t130}}
\subfigure[Subfigure 2 list of figures text][$t = 18.09$]{
\includegraphics[width=0.3\textwidth]{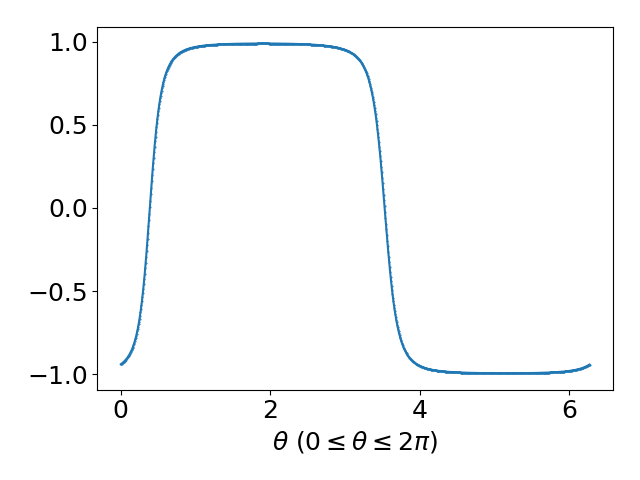}
\label{fig:TiltStep_t180}}
\caption{$\frac{e^{U-\eta}v_{1}}{\sqrt{1+e^{2(U-\eta)}v_{1}^{2}}}$ at various times. $N=1000$, $K=0.1$.}
\label{fig:tilt_step}
\end{figure}

The density gradient is, by definition, $\frac{\del_{\theta}\rho}{\rho}$. In terms of the re-scaled density \eqref{eqn:rhotdef}, it is given by
\begin{align*}
    \frac{\del_{\theta}\rho}{\rho} = \frac{\del_{\theta}\rhot}{\rhot}.
\end{align*}
As in \cite{BMO:2023}, we observe that the density gradient develops steep gradients and blows up as $t\rightarrow\infty$ for $K\in[0,1/3)$ if the initial spatial velocity vanishes at at least one point, as shown in Figure \ref{fig:density_gradient}. On the other hand, for $K \in [1/3,1)$, we have not observed such fluid spikes, which is consistent with that stability result \cite{BeyerOliynyk:2023}. This behaviour is shown in Figure \ref{fig:density_gradient_035}. The tilt instability is due to the fluid asymptotically approaching two different null vectors. This behaviour is particularly apparent when considering the fluid vector in an orthonormal basis. Using the metric \eqref{eqn:gowdymetricA}, we obtain the orthonormal frame vectors
\begin{align*}
    e_{0} = e^{U-\eta-\alpha}\del_{t}, \;\; e_{1} = e^{U-\eta}\del_{\theta},
\end{align*}
cf.\ \eqref{eq:specialorthframe}, for the special case that the metric is \eqref{eq:asymptoticFLRWmetric}.
Following the heuristics from Section~\ref{sec:asympbehav}, the variable that corresponds to $\beta$ is 
\begin{align*}
    \frac{e^{U-\eta}v_{1}}{\sqrt{1+e^{2(U-\eta)}v_{1}^{2}}}.
\end{align*}
As the fluid velocity approaches the two null vectors (with opposite tilts), we expect this quantity should approach a step function. This behaviour is confirmed in Figure \ref{fig:tilt_step}.

\FloatBarrier

\section{Large Perturbations: Fluid and Gravitational Spikes}
\label{sec:BigDataSpikes}
\subsection{Fluid Spikes for Large Initial Data}
\begin{figure}[htbp]
\centering
\subfigure[Subfigure 1 list of figures text][$t = 10.05$]{
\includegraphics[width=0.3\textwidth]{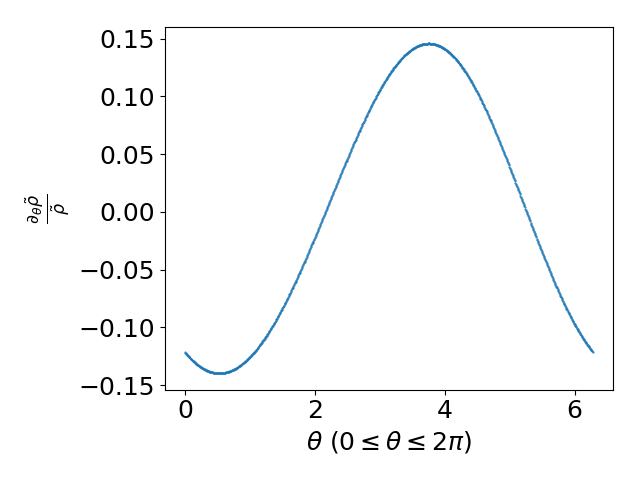}
\label{fig:subfigDC_t30_05}}
\subfigure[Subfigure 2 list of figures text][$t = 20.1$]{
\includegraphics[width=0.3\textwidth]{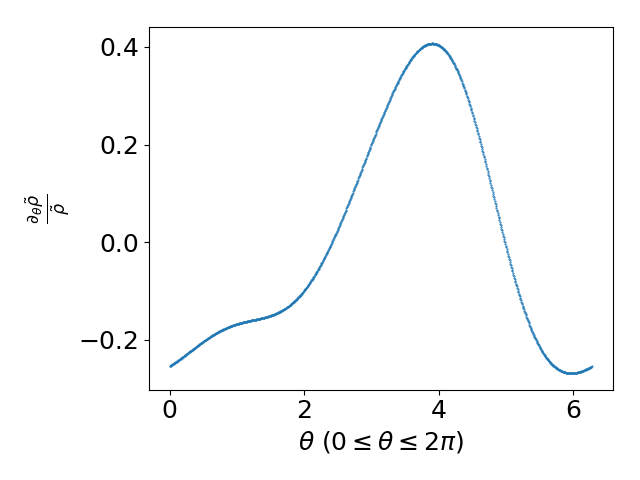}
\label{fig:subfigDC_t120_05}}
\subfigure[Subfigure 2 list of figures text][$t = 34.17$]{
\includegraphics[width=0.3\textwidth]{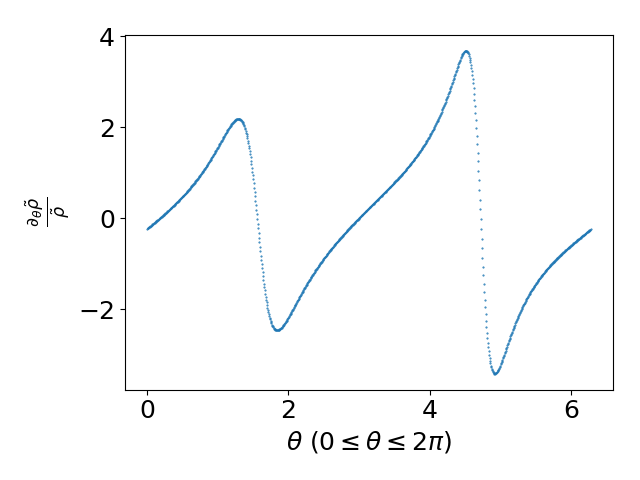}
\label{fig:subfigDC_t170_05}}
\caption{Density gradient $\frac{\del_{\theta}\rho}{\rho}$ at various times. $N=1000$, $K=0.5$, $a=b=c=d=k=0.01$, $f=0.5$.}
\label{fig:density_gradient_05}
\end{figure}

In the following, we will consider initial data with large values of the parameters, $a$, $b$, $c$, $d$, $f$, and $k$. As we increase the size of these parameters, our initial data becomes further away from that of the FLRW solution. We find that for suitably large initial data, fluid spikes form for \textit{all values of the parameter} $K$, shown in Figure \ref{fig:density_gradient_05}. 

As discussed in the introduction, this is consistent with the behaviour described in \cite{ColeyLim:2013} where a tilt instability was observed in inhomogeneous cosmological models for $0<K<1$. On first appearances, it may seem that these numerical results of this article  conflict with the stability of the FLRW big bang singularities for sounds speeds $1/3<K<1$ that was rigorously established in \cite{BeyerOliynyk:2023}. However, the stability established in \cite{BeyerOliynyk:2023} only holds for sufficiently small perturbations of FLRW solutions and, by choosing our initial data suitably large, we have exited the stable regime.

\subsection{Gravitational Spikes}
We now demonstrate the formation of gravitational spikes in numerical solutions to \eqref{eqn:EinsteinEuler1}-\eqref{eqn:EinsteinEuler2} for initial data of the form \eqref{eqn:numericalID} with the parameters $a$, $b$, $c$, $d$, $f$, and $k$ set as follows
\begin{equation}
\label{eqn:spikeID_parameters}
    a=b=c=d=k=0.01, \;\; f=0.5.
\end{equation}
We have been unable to observe spikes in the metric functions for values of $a$, $b$, $c$, $d$, $f$, and $k$ which are too close to zero, that is, near a FLRW solution. Following \cite{RendallWeaver:2001}, we confirm that spikes produced in our simulations are not coordinate artifacts by observing the behaviour of curvature invariants. If a curvature invariant shows spiky features at the same location as the metric functions, then the spikes are physical rather than gauge\footnote{Physical and gauge spikes are sometimes referred to as `true' and `false' spikes, respectively.}. 

\begin{figure}[htbp]
\centering
\subfigure[Subfigure 1 list of figures text][Density Gradient $\frac{\del_{\theta}\rho}{\rho}$]{
\includegraphics[width=0.3\textwidth]{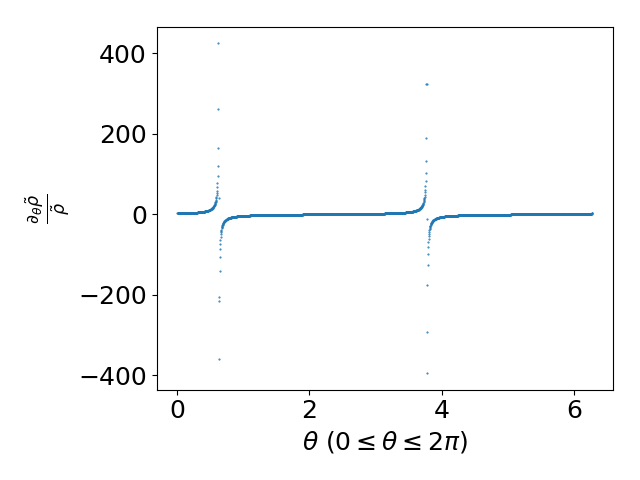}
\label{fig:subfig_spikesDG}}
\subfigure[Subfigure 2 list of figures text][$U$]{
\includegraphics[width=0.3\textwidth]{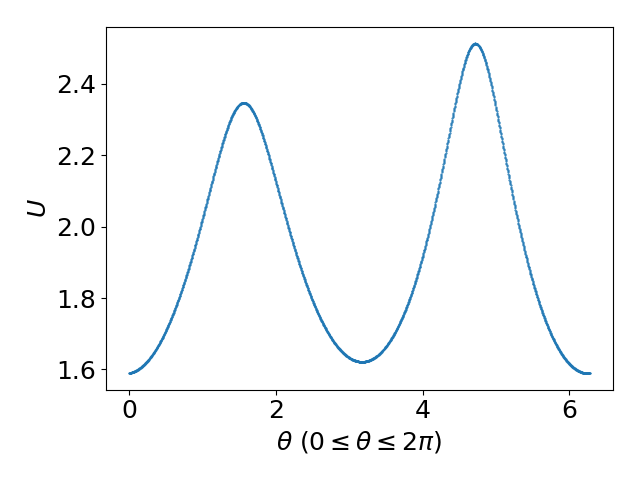}
\label{fig:subfig_spikesU}}
\subfigure[Subfigure 2 list of figures text][Ricci Scalar, $R$]{
\includegraphics[width=0.3\textwidth]{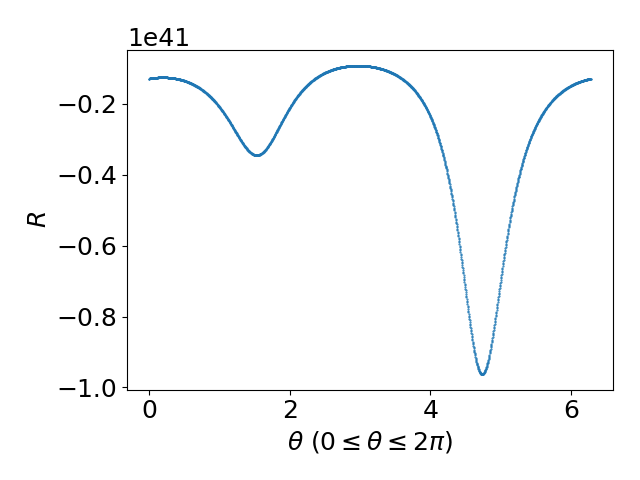}
\label{fig:subfig_spikesRicci}}
\caption{Density gradient, $U$, and the Ricci scalar $R$ at $t=25.62$. $N=2000$, $K=0.2$.}
\label{fig:Spikes_1}
\end{figure}

\begin{figure}[htbp]
\centering
\subfigure[Subfigure 1 list of figures text][$t = 0.753$]{
\includegraphics[width=0.45\textwidth]{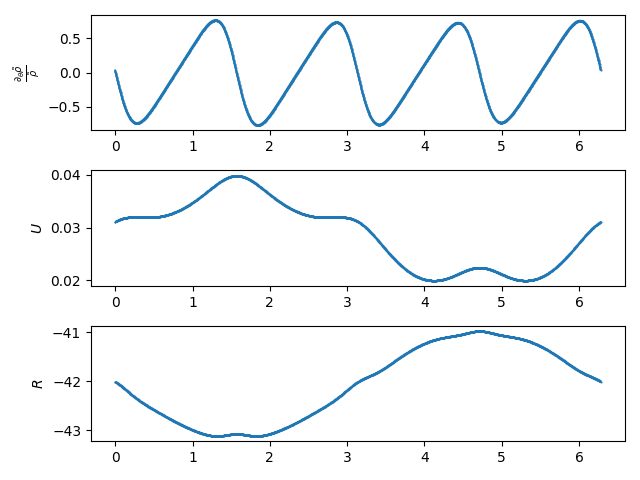}
\label{fig:subfigmultispike_t5}}
\subfigure[Subfigure 2 list of figures text][$t = 3.76$]{
\includegraphics[width=0.45\textwidth]{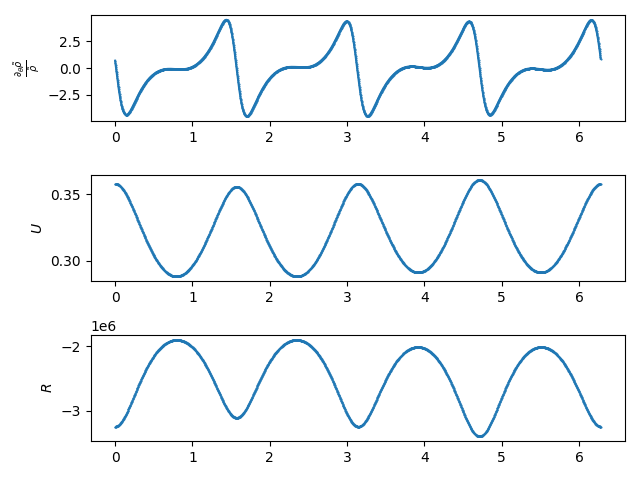}
\label{fig:subfigmultispike_t25}}
\subfigure[Subfigure 2 list of figures text][$t = 7.53$]{
\includegraphics[width=0.45\textwidth]{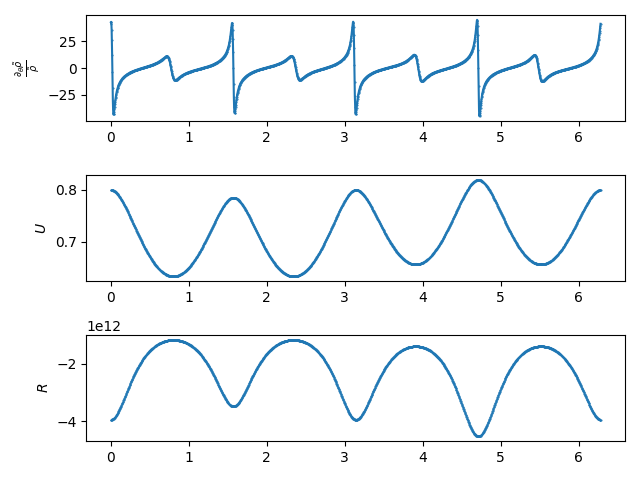}
\label{fig:subfigmultispike_t50}}
\subfigure[Subfigure 2 list of figures text][$t = 22.61$]{
\includegraphics[width=0.45\textwidth]{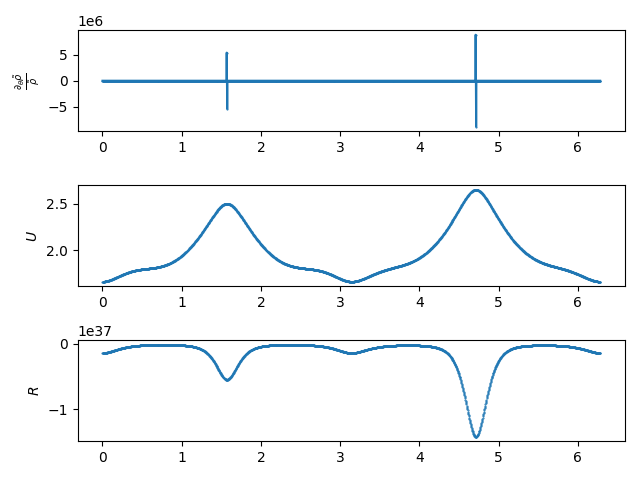}
\label{fig:subfigmultispike_t150}}
\caption{Density gradient $\frac{\del_{\theta}\rho}{\rho}$ (top), $U$ (middle), and the Ricci scalar $R$ (bottom) at various times. $N=2000$, $K=0.1$.}
\label{fig:multi_spikes}
\end{figure}

For the choice of parameters \eqref{eqn:spikeID_parameters}, we observe in our numerical simulations that spikes form at the same location in the fractional density gradient, the metric function $U$, and the Ricci scalar $R$ as shown in Figure \ref{fig:Spikes_1}. 
A natural question is whether the fluid spikes in the density gradient are related to the gravitational spikes in $U$ and the Ricci scalar. To test this, we consider initial data for which the fluid velocity crosses zero more than twice. In particular we modify our initial data \eqref{eqn:numericalID} by taking $\alpha$ and $v_{1}$ to be
\begin{equation}
\begin{aligned}
\label{eqn:numericalID_multiplespikes}
\mathring{v_{1}} &= a\sin(n\theta), \\
    \mathring{\alpha} &= -\log\Big(-\frac{a}{n}\cos(n\theta)+\sqrt{\frac{4}{3(K+1)}+2}\Big),
\end{aligned}
\end{equation}
where $n$ is an arbitrary positive integer which determines the number of times $v_{1}$ crosses zero. In practice, the value of $n$  corresponds to the number of fluid spikes that initially form in the fractional density gradient. For initial data of the form \eqref{eqn:numericalID_multiplespikes}, we observe that the spikes in the density gradient form first, followed by gravitational spikes in $U$ and the Ricci scalar at the same locations. At late times, we always observe two of the fractional density gradient spikes grow more rapidly than the others, which overwhelms the resolution of our simulations. These larger spikes dominate our plots at late times (i.e.\ near the big bang singularity), which we suspect masks `small' scale features that are related to the other smaller fractional density gradient spikes. This behaviour is demonstrated in Figures \ref{fig:subfigmultispike_t5} - \ref{fig:subfigmultispike_t150}. In particular, it is clear from Figure \ref{fig:subfigmultispike_t150} that our resolution is insufficient to capture any small scale features of the density gradient near the big bang singularity. 
To generate these plots, we have used the following choice of parameters 
\begin{equation}
\label{eqn:spikeID_parameters}
    b=c=d=k=0.01, \;\; a=0.1, \;\; f=0.5, \;\; \text{and} \;\; n =4,
\end{equation}
in our initial data.
It should be noted that while increasing the size of $a$ is not necessary to generate multiple fluid spikes, it makes the smaller spikes more apparent before they are overwhelmed by the dominant spikes at late times.

\section{Discussion}
In this article, we have numerically simulated Gowdy-symmetric, nonlinear perturbations of FLRW solutions to the Einstein-Euler-scalar field equations over the full sound speed parameter range $0\leq K\leq 1$ in the contracting direction. For $K\in (1/3,1]$, we observe numerically that sufficiently small perturbations of FLRW solutions are stable towards the past (contracting direction) and terminate in a spacelike big bang singularity in agreement with the analytic results obtained in \cite{BeyerOliynyk:2023, RodnianskiSpeck:2018c}. For these solutions, all suitably normalised gravitational and matter fields converge monotonically to limits on the big bang singularity.  We also observe similar stable behaviour for $K=1/3$, but as discussed above, a more thorough investigation is required to be confident that we are integrating long enough to resolve the asymptotic behaviour of solutions. On the other hand, for $K\in [0,1/3)$, we observe numerically that small perturbations of the FLRW solutions for which the spatial fluid velocity vanishes somewhere on the initial hypersurface are unstable towards the past. These solutions still terminate in the past at a spacelike big bang singularity, but now the fluid develops a fluid tilt-instability that manifests as sharp features (spikes) that develop in the fractional density gradient  $\frac{\del_{\theta}\rho}{\rho}$ and ultimately lead to blow-up of this quantity at
at finitely many spatial points on the big bang singularity. Interestingly, a similar fluid tilt-instability in the expanding direction (to the future) was predicted by Rendall \cite{Rendall:2004} and observed numerically in  \cite{BMO:2023}. We have also observed that for initial data suitably far away from that of the FLRW solution, gravitational spikes form in the metric functions as well as fluid spikes for all $K\in [0,1]$. While it appears from the numerical simulations that the gravitational spikes are induced by the fluid spikes, more investigation is required to understand the precise relationship. We plan on investigating this further in future work. The results of this article suggest several interesting topics for future research. The obvious first step is to remove the Gowdy symmetry assumption and study the fluid tilt-instability that develops in small perturbations of FLRW solutions for $K\in [0,1/3)$ without any additional symmetry assumptions. Additionally, it would be interesting to further study the connection between dynamics of the Einstein-Euler system with positive cosmological constant towards the future with the Einstein-Euler-scalar field system towards the past.

\bibliographystyle{amsplain}
\bibliography{refs}

\end{document}